\documentclass{aa}
\usepackage{graphicx}
\usepackage{txfonts}
\usepackage{natbib}
\usepackage{mathrsfs}
\usepackage{longtable}
\usepackage[shortlabels]{enumitem}
\usepackage{siunitx}
\usepackage{epstopdf}
\usepackage[]{hyperref}	% Hyperlinks
\hypersetup{colorlinks=true,linkcolor=blue,citecolor=blue,filecolor=blue,urlcolor=blue}
\usepackage{comment}
\usepackage{txfonts}
\usepackage{url}

\usepackage{overpic}
\begin{document}
\title{APOGEE DR16: a multi-zone chemical evolution model for the Galactic disc  based on  MCMC methods}
\author {E. Spitoni\inst{1} \thanks {email to: spitoni@phys.au.dk}
  \and K. Verma\inst{1} \and
  V. Silva  Aguirre\inst{1}  \and F. Vincenzo  \inst{2,3}   \and F. Matteucci  \inst{4,5,6} \and \\
  B. Vai\v{c}ekauskait\.{e}   \inst{7,1}\and M. Palla  \inst{4,5,6}
  \and   V. Grisoni  \inst{8,5} \and F. Calura  \inst{9}}
\institute{Stellar Astrophysics Centre, Department of Physics and
  Astronomy, Aarhus University, Ny Munkegade 120, DK-8000 Aarhus C,
  Denmark \and Center for Cosmology and AstroParticle Physics, The Ohio State University, 191 West Woodruff Avenue, Columbus, OH 43210, USA \and Department of Astronomy, The Ohio State University, 140 West 18th Avenue, Columbus, OH 43210, USA \and Dipartimento di Fisica, Sezione di Astronomia,  Universit\`a di Trieste, via
  G.B. Tiepolo 11, I-34131, Trieste, Italy \and I.N.A.F. - Osservatorio
  Astronomico di Trieste, via G.B. Tiepolo 11, I-34131, Trieste,
  Italy \and I.N.F.N. -  Sezione di Trieste, Via Valerio 2, I-34100
  Trieste 
  \and Technology University Dublin, School of Physics and Clinical \& Optometric Sciences, Kevin Street, Saint Peter's, Dublin 2, D08 X622, Ireland
  \and S.I.S.S.A. - International School for Advanced Studies, via Bonomea 265, I-34136, Trieste, Italy
\and I.N.A.F. - Osservatorio Astronomico di Bologna, Via Gobetti 93/3, 40129 Bologna, Italy}
\date{Received xxxx / Accepted xxxx}

\abstract
{The analysis of the APOGEE DR16 data suggests the existence of a clear distinction between two sequences of disc stars at different Galactocentric distances in the [$\alpha$/Fe] vs. [Fe/H] abundance ratio space: 
the so-called high-$\alpha$  sequence, classically associated to an old population of stars in the thick disc with high average [$\alpha$ /Fe], and the low-$\alpha$  sequence, which mostly comprises relatively young stars in the thin disc with low average [$\alpha$/Fe].
}
%%%
%%%
{We aim to constrain a multi-zone  two-infall chemical evolution model designed for regions at different Galactocentric distances  using measured chemical abundances  from the  APOGEE DR16 sample.}
%%%
%%%
{We perform a Bayesian analysis based on  a Markov Chain Monte Carlo method to fit our  multi-zone two-infall chemical evolution model to the APOGEE DR16 data.}
%%%
%%%
{
An inside-out formation of the Galaxy disc naturally emerges from the best fit of our two-infall chemical-evolution model to APOGEE-DR16: inner  Galactic regions are assembled on shorter time-scales compared to the external ones.
In the outer disc  (with radii  $R>6$ kpc), the chemical dilution due to a late  accretion event of gas with primordial chemical composition is the main driver of  the [Mg/Fe] vs. [Fe/H] abundance pattern in the low-$\alpha$ sequence.
In the inner disc, in the framework of the two-infall model, we confirm the presence of an enriched gas infall in the low-$\alpha$ phase as suggested by  chemo-dynamical models. Our Bayesian analysis of the recent APOGEE DR16 data suggests a significant delay time,  ranging  from $\sim$3.0 to 4.7 Gyr, between the first and second gas infall events   for  all  the analyzed  Galactocentric  regions. 
The best fit model reproduces several observational constraints such as: (i) the present-day stellar and gas surface density profiles; (ii) the present-day abundance gradients; (iii) the star formation rate profile;  and (iv) the solar abundance values.
}
 {Our  results  propose a clear interpretation of the [Mg/Fe] vs. [Fe/H] relations along the Galactic discs.  The signatures of a delayed  gas-rich merger which gives  rise to  a hiatus in the star formation  history of the Galaxy are impressed in   the  [Mg/Fe] vs. [Fe/H] relation, determining how  the  low-$\alpha$ stars are distributed in the abundance space at different  Galactocentric distances, in agreement with the finding of  recent chemo-dynamical simulations.}


\titlerunning{Multi-zone chemical evolution model based on MCMC methods}

\authorrunning{Spitoni et al.}

\maketitle
\section{Introduction}

Our understanding of the formation and evolution of our Galaxy disc is essentially based on the  study and interpretation of signatures
imprinted in resolved stellar populations, such as 
their chemical and kinematic properties as traced by  large surveys and observational campaigns. 
The current synergy between  the Apache Point Observatory Galactic Evolution
Experiment project (APOGEE; \citealt{Majewski:2017ip}; 
in particular the latest data release DR16, \citealt{Ahumada2019})
and the Gaia mission \citep[DR2;][]{gaia2_2018}, 
offers  an unparalleled opportunity 
to simultaneously rely upon 
accurate spectroscopic and kinematic properties 
to constrain models of Galactic chemical evolution.

The analysis of the APOGEE DR16 data  \citep{Ahumada2019,queiroz2020} suggests the existence of a clear distinction between two sequences of disc stars in the [$\alpha$/Fe] vs. [Fe/H] abundance ratio space: the so-called high-$\alpha$  and low-$\alpha$ sequences.
This dichotomy  in the chemical  abundance ratio space  has been also confirmed by  the Gaia-ESO  survey \citep[e.g.,][]{RecioBlanco:2014dd,RojasArriagada:2016eq,RojasArriagada:2017ka}, the AMBRE project \citep{Mikolaitis:2017gd}, and the GALAH survey \citep{buder2019}.

By analyzing  the  APOKASC (APOGEE+ Kepler Asteroseismology Science Consortium) sample for the solar neighborhood, \citet{victor2018} pointed out that the two sequences are characterized by two different ages: the high-$\alpha$ stars have ages of $\sim$11\, Gyr, while the low-$\alpha$  sequence peaks at $\sim$2 Gyr. 
Several theoretical models of the Galactic discs evolution suggested that the  bimodality may be strictly connected to a delayed gas  accretion episode of primordial composition. By revising the classical two-infall chemical evolution model by \citet{chiappini1997} and \citet{grisoni2017}, \citet{spitoni2019, spitoni2020} have shown that a significant delay ranging from 4.5 to 5.5\,Gyr between two consecutive episodes of 
gas accretion is needed to explain the dichotomy in the local APOKASC sample \citep{victor2018}.
In particular, they predicted that the  star formation
rate (SFR) has a minimum at an age of $\sim \! 8$\,Gyr. A similar quenching of
star formation around the age of 8\,Gyr was derived by \citet{snaith2015}  using the chemical abundances of 
 \citet{adi2012}  and the isochrone ages of  \citet{haywood2013} for 
solar-type stars. Moreover,  \citet{mor2019}  found  indications  of a  possible double-peaked star formation 
history  with a minimum around 6\,Gyr,
from the analysis of $Gaia$ colour-magnitude diagrams. 
  By analyzing the High Accuracy Radial velocity Planet Searcher (HARPS) spectra  of local solar twin stars, \citet{nissen2020} found that the age-metallicity distribution has two distinct populations with a clear age dissection. The authors suggested that these two sequences  may be interpreted as evidence of two episodes of accretion of gas onto the Galactic disc with  quenching of star formation in between them, in agreement with the scenario proposed by \citet{spitoni2019,spitoni2020}. 

In a cosmological framework, the existence of a double sequence was predicted for the first time by \citet{calura2009} who, by means of a semi-analytic model based on the extended Press and Schechter \citep{bond1991},  modelled in post-processing the abundance pattern of a sample of model galaxies selected as Milky Way analogues. 
A second accretion phase after a prolonged period with a quenched star formation  has been 
suggested by the dynamical models of \citet{noguchi2018} in which a  
 first infall episode rapidly builds up  the high-$\alpha$ sequence,
 but then the star formation is starved from the lack of gas supply from the intergalactic medium (IGM)
 until the shock-heated gas in the Galactic dark matter halo has radiatively cools down and is accreted by the Galaxy giving rise to a delayed second gas  infall episode.  In this framework, \citet{noguchi2018}  found that the  SFR of the Galactic disc is characterised by two distinct peaks separated by  $\sim 5$\, Gyr.
Moreover, the AURIGA simulations presented 
by \citet{grand2018} clearly point out that a bimodal distribution in the [Fe/H]-[$\alpha$/Fe] plane may be a 
consequence of a significantly lowered gas accretion rate at ages between 6 and 9 Gyr.
In the framework of cosmological hydrodynamic simulations of Milky Way like galaxies, \citet{buck2020}  for example found that a 
dichotomy in the  $\alpha$-sequence is a generic consequence of a gas-rich merger occurred at a certain epoch in the evolution of the Galaxy, which destabilized the gaseous disc at high redshift.

\begin{figure*}
\begin{centering}
\includegraphics[scale=0.53]{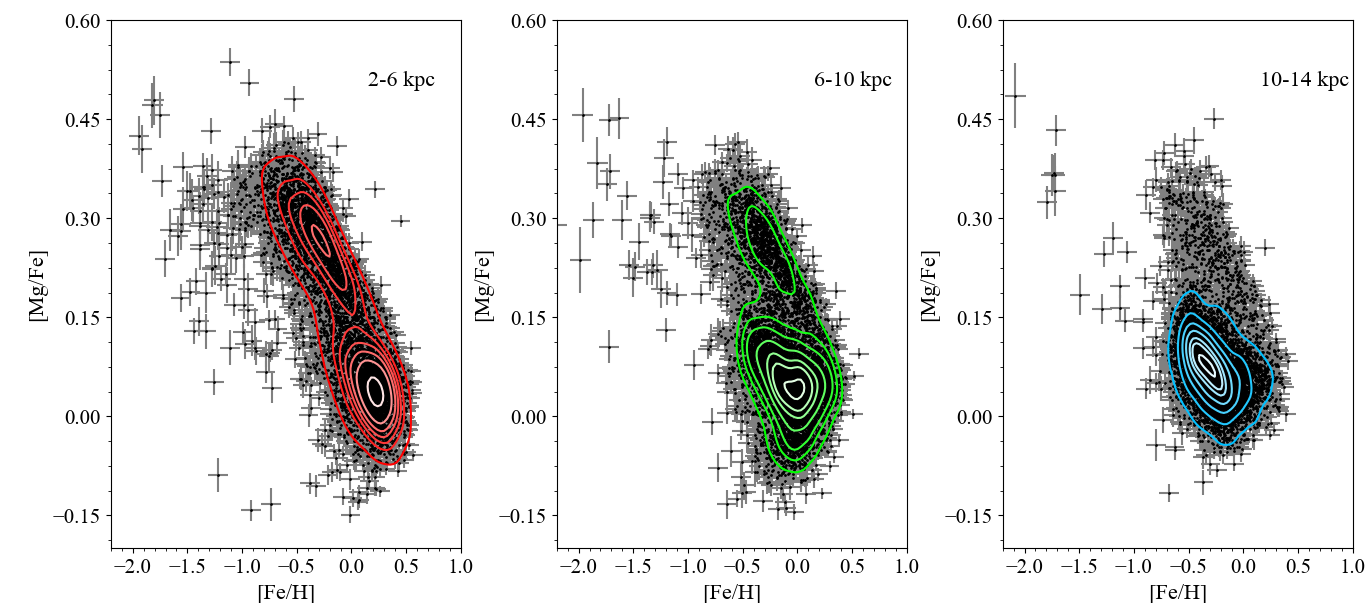}
\caption{Observed stellar [Mg/Fe] vs. [Fe/H] abundance ratios from APOGEE DR16 \citep{Ahumada2019} for three  bins of different  Galactocentric distances. The regions are 4 kpc-wide centered at 4 kpc (left panel), 8 kpc (middle panel) and 12 kpc (right panel), respectively. 
The contour lines enclose fractions of 0.90, 0.75, 0.60, 0.45, 0.30, 0.20, 0.05 of the total number of observed stars.
 Details on the data selection are reported in the text.  }
\label{apogee_data}
\end{centering}
\end{figure*}

 The significant delay in the  two-infall model of \citet{spitoni2019,spitoni2020}
has been  discussed by \citet{vincenzo2019} in the context  of the stellar system accreted by the Galactic halo, AKA  Gaia-Enceladus  \citep{helmi2018,koppelman2019}.  It was proposed that the mechanism which quenched the Milky Way star formation at high redshift  was a major merger event with a satellite like Enceladus (by heating 
up the gas in the dark matter halo).  This proposed scenario is in agreement with the recent  \citet{chaplin2020} study. They constrained the merging time with the observations of the very bright, naked-eye star $\nu$ Indi, finding at 68\% confidence that the earliest the merger could have started was 11.6 Gyr ago.

 As  outlined   by \citet{hayden2015} and \citet{queiroz2020}, the two sequences  in the [$\alpha$/Fe] vs. [Fe/H] abundance ratio relation  from  APOGEE   have different features and trends throughout the Galactic disc.  While the low-$\alpha$  sequence is 
 distributed at increasingly lower metallicity towards the outer disc, it is found at super-solar values in the inner disc.
 Moreover, it is worth noticing that the ratio between the number of low-$\alpha$ and  high-$\alpha$  stars increases when moving from the inner to the outer Galactic disc.
 Hence, the formation of the low-$\alpha$ sequence in  the entire Galaxy seems to be more complex than a simple sequential process as assumed in the   model of \citet{spitoni2020} for the solar vicinity and may be driven by different physical processes.   
 For instance, in the cosmological simulations presented by  \citet{VINTERGATANI2020},  \citet{VINTERGATANII2020} and \citet{VINTERGATANIII2020}, they concluded that the low-$\alpha$ sequence has been assembled through different physical process that interplay together in the whole disc.
Two distinct channels of gas infall fuel the Galactic disc; a chemically enriched  gas accretion event (by outflows from a  massive galaxy with $\sim$ 1/3 of the Galactic mass at the time of the interaction) feeds the inner Galactic region, whereas a different one fuels the outer gas disc, which is inclined with respect to the main Galactic plane and has significantly poorer chemical content. 
However, their predicted low-$\alpha$ sequence is shifted towards larger 
[$\alpha$/Fe] values than the APOGEE sample by  $\sim$ 0.3 dex.

Recently, \citet{palla2020} presented a revised Galactic chemical evolution model for  the  disc formation based on  the two-infall scenario in order to reproduce  the observed [Mg/Fe] vs. [Fe/H] of APOGEE \citep{hayden2015} at different Galactocentric distances. \citet{palla2020} proposed that a delay  of $t_{max}$= 3.25 Gyr  between the two gas infall events invoking an  enriched gas infall to properly reproduce the inner disc  [Mg/Fe] vs. [Fe/H] abundance ratio.
\citet{Khoperskov2020} found that in  the infalling gas during inner thin disc formation phase is not primordial because the gaseous halo has been significantly polluted during the formation of the  thick disc, providing a tight connection between chemical abundance patterns in the two Galactic disc components.

In this article, we  present  a  multi-zone two-infall chemical evolution model with the aim to extend the results of \citet{spitoni2019,spitoni2020} for the solar vicinity to the whole disc.
We quantitatively infer the free parameters by fitting the APOGEE DR16 \citep{Ahumada2019}  abundance ratios  at  different Galactocentric distances using a Bayesian technique based on Markov Chain Monte Carlo (MCMC) methods. 
The Bayesian analysis is now being widely used in testing  the Galactic chemical evolution models 
\citep[see e.g.][]{cote2017,rybi2017,philcox2018,frankel2018,belfiore2019,spitoni2020}.
In fact, thanks to the wealth of information from large
surveys, large datasets are currently being exploited 
by means of statistic methods to constrain the parameters of Galactic chemical evolution models.

The paper is organised as follows. In Section \ref{s:apogee},   the observational data used in the Bayesian analysis  are presented. In Section \ref{s:chemmod}, we present the main characteristics of the  multi-zone chemical evolution model adopted in this work, and describe the fitting method.  
In Section \ref{results}, we present our results, and finally in Section \ref{conc}, we draw our conclusions.

\section{The APOGEE DR16 sample}\label{s:apogee}
  Here, following \citet{spitoni2020},  we use a Bayesian framework based on MCMC methods to fit a multi-zone Galactic chemical evolution model to the observed chemical abundances for  Mg and Fe provided by APOGEE DR16 \citep{Ahumada2019} and related Galactocentric distances and vertical heights above the Galactic plane, as found   by  the $Gaia$ mission \citep[DR2;][]{gaia2_2018}. 
  
   Different methods have been introduced to compute proper Galactocentric distances. \citet{luri2018} highlighted that the estimation of distances from parallaxes has to be addressed as a fully Bayesian inference problem, as shown in  \citealt{BJ2018} for $Gaia$ data. Moreover, in the Bayesian framework \citet{queiroz2018} presented spectro-photometric  distances estimated with the StarHorse tool which, in the case of APOGEE DR16, are derived from a set of photometric bands, APOGEE spectra and $Gaia$ parallaxes.
In this paper, we adopt the Galactocentric distances computed  by \citet{leung2019} and reported  in the astroNN \footnote{\href{https://data.sdss.org/sas/dr16/apogee/vac/apogee-astronn}{https://data.sdss.org/sas/dr16/apogee/vac/apogee-astronn}} 
catalogue  for APOGEE DR16 stars. 
 To obtain precise distances for distant stars, \citet{leung2019} designed a deep neural network, and trained it using precisely measured parallaxes of nearby stars in common between $Gaia$ and APOGEE to determine the spectro-photometric distances for APOGEE stars. They included a flexible model to calibrate parallax zero-point biases in $Gaia$ DR2  in order to avoid the propagation of systematic uncertainties present in the training data set to the inferred distances. On top of the versatility of the neural network, they employed a robust way of Bayesian deep learning that takes data uncertainties in the training set into account and also estimates uncertainties in predictions made with the neural network using the drop out variational inference. One major limitation of the method is the size of the training set it rests upon.  Fortunately, the amount of data to train their algorithm will increase in the future thanks to new APOGEE and $Gaia$ data releases and other spectroscopic surveys such as GALAH.

The Galactocentric positions and velocities used in the computation of the orbital properties are calculated assuming that the  Galactocentric distance $R_{\odot}$ from  the Sun  to the Galactic center is 8.125 kpc \citep{abuter2018} and located 20.8 pc above the Galactic midplane \citep{bennett2019}. 
  
  Stars that are part of the Galactic disc have been chosen with the same quality cuts suggested in  \citet{weinberg2019} assuming signal-to-noise ratio $(S/N) >80$,  logarithm of surface gravity  between $1.0< \log g<2.0$ and vertical height  $\lvert z \rvert<$ 1 kpc to have cleaner trends in the [$\alpha$/Fe] vs [Fe/H] abundance patterns at different radii for the MCMC fitting, since we adopt a one-zone chemical evolution model for each radial range of Galactocentric distances, which does not account for the small displacements observed in the APOGEE abundance distributions of the high-$\alpha$ and low-$\alpha$ sequences at high $|z|$ \citep{hayden2015}. 
  
  The uncertainties in metallicity reported in this compilation correspond to the internal precision, which are of the order of $\sim$ 0.01 dex. In the effort to better estimate systematic uncertainties, we added, as in \citet{victor2018}, the median difference between APOGEE results for clusters and the standard literature values as reported in Table 3 of \citet{tayar2017} ($\sim 0.09$ dex) in quadrature. 
  
We have split the data into  3 concentric annular Galactic regions, each 4 kpc-wide, spanning the range between 2  and 14 kpc. 
In   Fig. \ref{apogee_data},  it is clear that the further out stars   (right panel, 10-14 kpc region)  preferentially populate the low-$\alpha$ sequence in the [Mg/Fe] vs. [Fe/H] relation, and few stars are located in the high-$\alpha$ population. Moreover, the more  the regions are external,  the more the  locus of the low-$\alpha$ sequence is shifted  towards lower metallicity. 
On the contrary, in the annular region enclosed between 2 and 6 kpc  (left panel  of Fig. \ref{apogee_data}),  the low-$\alpha$ phase peaks at  super-solar metallicity, however a  clear bimodality is still evident as highlighted by the isodensity contours.

 We did not consider innermost regions with $R<2$ kpc, because there the Galactic bulge  is the dominant component and different Galactic chemical assumptions and prescriptions need to be applied  \citep{matteucci2019,matteucci2020,griffith2020}.
The recent analysis of large samples from APOGEE DR16 data \citep{queiroz2020, rojas2020} suggested that bulge structure could  extend up to Galactocentric distances of  $3-3.5$ kpc. Hence, a partial contamination of bulge stars could in principle affect the region enclosed between 2 and 6 kpc. However, we checked that   stars with Galactocentric distances  selected in the range between  2 and 6 kpc   give place to  an almost identical  distribution  in the [Mg/Fe] vs. [Fe/H] abundance ratio space as the ones enclosed in  a region between 3 and 6 kpc.

The numbers of stars in the considered  different annular regions are the following ones: 7440 in the zone centered at 4 kpc, 9169 in the one at 8 kpc, and 10081 in the outermost region centered at 12 kpc.
We believe that in  these three zones  the main trends of the Galactic disc in  terms of the  [Mg/Fe] vs. [Fe/H]  abundance  as a function of the Galactocentric distance are imprinted .  We have checked that the 4 kpc-wide region centered at 16 kpc has only 882 stars, almost all of which are in the low-$\alpha$ sequence.

\section{Multi-zone chemical evolution model for the  Galactic disc  }\label{s:chemmod}
In this Section we present the main assumptions  and characteristics of the multi-zone  chemical evolution model which extends the previous ones introduced by \citet{spitoni2019,spitoni2020} for the solar neighborhood.
After  a brief explanation  of   the  reasons for  neglecting  stellar migration effects,  we describe the adopted  MCMC methods. 
\subsection{Chemical evolution model prescriptions}\label{presc}

We extend the chemical evolution model designed for the solar neighborhood presented  by \citet{spitoni2019,spitoni2020} to different Galactocentric regions centered at  4 kpc, 8 kpc and 12 kpc. 
Retaining the assumption that the Milky Way disc has been formed by two distinct accretion
episodes of gas,  we assume that  the gas infall rate is a function of the Galactic distance $R$ \citep{chiappini2001,grisoni2018,spitoni2009,spitoni2D2018} and     is expressed by the following expression,

\begin{eqnarray}\nonumber
\mathcal{I}_i(t,R)&=& \mathcal{X}_{1,i}(R) \mathcal{N}_1(R) \, e^{-t/ \tau_{1}(R)}+ \\
&& +\theta(t-t_{{\rm max}, \,R}) \, \mathcal{X}_{2,i}(R) \, \mathcal{N}_2(R) \, e^{-(t-t_{{\rm max}, \,R})/ \tau_{2}(R)},
\label{infall}
 \end{eqnarray}
where $\tau_{1}(R)$ and $\tau_{2}(R)$  are the time-scales of gas accretion for the formation of the
high-$\alpha$ and low-$\alpha$ disc phase, respectively. 
The  quantity $\theta$ in the eq. (\ref{infall}) is the Heaviside step function. 
  $\mathcal{X}_{1,i}(R)$ and $\mathcal{X}_{2,i}(R)$ are the abundance by mass of the element $i$ in the
infalling gas for the first and second gas infall, whereas $t_{{\rm max}, \,R}$ is the time of the maximum infall rate on the  second accretion episode, i.e. it indicates the delay of 
the beginning of the second infall.  \citet{spitoni2019} underlined the importance of a consistent delay  of $t_{{\rm max}} \sim$ 4 Gyr in the solar vicinity (defined as an annular region 2 kpc-wide centered at $R=8$ kpc)  in order to properly reproduce in the solar neighborhood  the stellar abundances and asteroseismic ages of the APOKASC data sample \citep{victor2018}. This finding was confirmed later by   \citet{spitoni2020} using  a Bayesian analysis based on MCMC methods (see also \citealt{palla2020}).

Finally, the coefficients $\mathcal{N}_1(R) $ and $\mathcal{N}_2(R)$ are obtained by imposing a
 fit to the observed current total surface mass density at different radii $R$ with  the following relations:

\begin{equation}
\mathcal{N}_1(R) =\frac{\sigma_1(R)}{\tau_{1}(R) \left(1- e^{-t_G/\tau_{1}(R)}\right)},
\label{S1}
\end{equation}

\begin{equation}
\mathcal{N}_2(R) =\frac{\sigma_2(R)}{\tau_{2} (R)\left(1-
 e^{-(t_G-t_{{\rm max}, \, R})/\tau_{2}(R)} \right)},
\label{S2}
\end{equation}
where $\sigma_1(R)$ and $\sigma_2(R)$ are the present-day total surface mass density of the high-$\alpha$ and low-$\alpha$ sequence stars,
respectively, and $t_G$ is the age of the Galaxy.

Following \citet{spitoni2020}, we  use the value of total surface density (sum of high-$\alpha$ and low-$\alpha$) in the solar neighborhood of 47.1 $\pm$ 3.4 M$_{\odot} \mbox{ pc}^{-2}$ as provided by \citet{mckee2015}.
In  \citet{spitoni2020}  it was assumed that in the solar neighborhood  the total surface mass densities ($\sigma_{tot, \, \odot} =  \sigma_{1, \, \odot}+\sigma_{2, \, \odot}$) is constant, as given by  \citet{mckee2015}.
The  present-day total surface mass density at a certain Galactocentric distance $R$ can be written as
\begin{equation} 
 \sigma_{tot}(R)=  \sigma_{tot, \, \odot} \, e^{-(R-R_{\odot})/R_d},
\label{tot_exp}
\end{equation}
after having imposed that the total mass declines with the radius through an exponential law and the scale-length of the disc is $R_d=3.5$ kpc.
In contrast with  \citet{palla2020}  where different scale-lengths for the thick and thin disc phases were tested and  assumed (see their Fig. 6),  here we consider  the ratio between the surface gas densities  as  a free parameter of the model.

Recalling that $\sigma_2/\sigma_1$ is the ratio between the low-$\alpha$  and high-$\alpha$,  the values of the present-day total surface mass densities $\sigma_2(R)$ and  $\sigma_1(R)$ to insert in eqs. (\ref{S2}) and (\ref{S1}) are the following ones:
\begin{equation} 
 \sigma_2(R)= \frac{ \sigma_{tot}(R)} { 1+\left( \frac{\sigma_2}{\sigma_1}  \Big|_R \right)^{-1}     },
\label{newD}
\end{equation}
\begin{equation} 
\sigma_1(R) =\sigma_{tot}(R)- \sigma_2(R).
\end{equation}
The SFR is expressed as the \citet{kenni1998} law,
\begin{equation}
\psi(t,R)\propto \nu(t, R) \sigma_{g}(t,R)^{k},
\label{k1}
\end{equation}
 where $\sigma_g$ is the gas surface
 density and $k = 1.5$ is the exponent. 
The quantity $\nu(t,R)$ is the star formation efficiency (SFE). 
Motivated by the theory of star formation induced by spiral density
waves in Galactic discs \citep{wyse1989}, we consider a
variable SFE as a function of the
Galactocentric distance  in the low-$\alpha$ phase. In several chemical evolution models \citep{colavitti2008,spitoni2015,grisoni2018,palla2020}  it has been claimed that  observed abundance gradients  in the Galactic disc  may be explained by assuming  higher SFE values in the inner regions than in the outer ones (along with the 'inside-out' formation scenario and   radial gas flows).

Moreover,  different infall episodes  in principle could be characterized by different SFEs.
In fact, in the classical two-infall model
\citep{chiappini2001,grisoni2017,grisoni2019,grisoni2020,spitoni2020} the SFEs associated to the high-$\alpha$ and
low-$\alpha$ sequences are  different: $\nu_1=2$ Gyr$^{-1}$ and
$\nu_2=1$ Gyr$^{-1}$ for the solar vicinity.

 We adopt the \citet{scalo1986}  initial stellar mass function (IMF), constant in
time and space.

Although an important ingredient of the \citet{Nidever:2014fj} chemical evolution model to reproduce the APOGEE data was the inclusion of Galactic winds proportional to the SFR coupled to a variable loading factor, in this paper we do not consider outflows. In fact, while studying the Galactic fountains originated by the explosions of Type II SNe in OB associations, \citet{melioli2008, melioli2009} and \citet{spitoni2008, spitoni2009} found that the ejected metals fall back close to the same Galactocentric region where they are delivered and thus do not modify significantly the chemical evolution of the disc as a whole. 

 As in \citet{spitoni2019, spitoni2020},    we adopt the same  nucleosynthesis prescriptions as proposed  by \citet{francois2004} for  Fe, Mg and Si.
  The authors artificially increased the Mg yields for massive stars from \citet{WW1995} to reproduce the solar Mg abundance.
 Mg yields from stars in the range 11-20 M$_ {\odot}$ have been increased by a factor of  7,
 whereas yields for stars with mass  $>20$ M$_ {\odot}$  are on average a factor $\sim$ 2 larger. 
 No modifications are required for the yields of  Fe, as computed for solar chemical composition. Concerning Si, only the yields of  very massive stars (M $>$ 40 M$_ {\odot}$) are increased by a factor of 2. 
 Concerning  Type Ia SNe, in order to preserve the observed [Mg/Fe] vs. [Fe/H] pattern, the yields of \citet{iwamoto1999} for Mg were increased by a factor of 5.

This set of yields has been widely used in the literature \citep{cescutti2007,spitoni2014,spitoni2015, spitoni2017,spitoni2D2018, mott2013, vincenzo2019} and turned out to be able to reproduce the main features of the solar neighbourhood.
We  adopt the photospheric values of \citet{asplund2005} as our solar reference abundances, in order  to be consistent with the APOGEE DR16 release.    

\subsection{Stellar migration and  Galactic chemical evolution }

The existence  of stellar radial migration is established beyond any doubts \citep[see, e.g.,][]{roskar2008,schoenrich2009MNRAS, loebmn2011, minchev2012, kubryk2013}. 
The main physical mechanisms responsible for stellar migration are churning and blurring  \citep[e.g.,][]{sellwood2002,minchev2011}, as well as the overlap of the spiral and bar resonances on the disc \citep{minchev2011}.
However, the real impact of radial migration on the chemical evolution of the Galactic disc  is still under debate.

 \citet{Nidever:2014fj} and \citet{sharma2020} presented chemical evolution models where the dichotomy in the abundance space is entirely explainable only in terms of stellar migration.  They concluded that the high-$\alpha$ disc has been built by migrator stars and gas of the thin disc.
 
 In particular, the model presented by \citet{sharma2020}  assumed empirical tracks for the evolution of [$\alpha$/Fe] and [Fe/H] as a function of time at different radii. These empirical age-metallicity relations may be inconsistent with their assumed empirical relations for the SFR as a function of time at different radii, which did not require a hiatus  in the star formation  between the high-$\alpha$ and low-$\alpha$ sequences, as predicted by the two-infall chemical evolution  model and chemo-dynamical  simulations.  Finally, we note that stellar migration in  \citet{sharma2020} follows a parametric diffusion approach \citep[e.g.,][]{schoenrich2009MNRAS}.

On the contrary,  the analysis of recent results from    chemo-dynamical simulations has raised important doubts  about the importance of stellar migration   in the evolution of chemical abundance ratios, such as [$\alpha$/Fe] vs. [Fe/H],  in the Galactic disc.
For instance, by means of a self-consistent
chemo-dynamical model for the Galactic disc evolution, \citet{Khoperskov2020} 
concluded that radial migration
 has a negligible effect on the [$\alpha$/Fe] vs. [Fe/H] distribution over time (the  distribution is slightly smoothed by migrators  from the inner and outer disc regions), suggesting the $\alpha$-dichotomy is strictly linked to  different star formation regimes over the Galaxy’s life.
 
Similar results are found by the cosmological simulation of \citet{vincenzo2020}. 
The authors concluded that the two main  gas accretion episodes occurred    0-2 and 5-7 Gyr ago, determinant for the rise of the double sequence  in the  [$\alpha$/Fe] vs. [Fe/H] plot.
In their Fig. 13, it is  remarkable   that the abundances in stars  with ages smaller than 8 Gyr in the solar neighborhood perfectly trace the gas phase abundances  in the same  region. They  concluded that the signature impressed in the chemical abundances of the stars may be  linked to infalling of primordial or poorly enriched gas.
Supported by the these results, we can safely assume  that the stellar migration did not  alter or affect much the   [Mg/Fe] vs. [Fe/H] evolution in our analysis which adopts  Galactic annular  regions  4 kpc-wide.

  Although   stellar migration has played an important role in  Galactic evolution, i.e. in  flattening of the radial metallicity profiles and affecting the [$\alpha$/Fe]-age relation of thin-disc stars \citep{vincenzo2020}, 
   we investigate a complementary scenario with respect to that proposed by \citet{sharma2020}, in which the radial variation in the   [$\alpha$/Fe] vs. [Fe/H] abundance   have been entirely caused  by the stellar migration.

\subsection{Fitting the data with MCMC methods}\label{fitting}

As in \citet{spitoni2020}, we use a Bayesian analysis based on Markov Chain Monte Carlo (MCMC) methods to find the best-fit chemical evolution models at different Galactocentric distances, $R$. Here,  we briefly recall the main assumptions and refer  the  reader to  \citet{spitoni2020} for  a more detailed description of the fitting method.

At a fixed $R$,  the set of observables is
${\bf x_R}=\{[\rm Mg/{\rm Fe}], \ [{\rm Fe}/{\rm H}]\}$ while the set of model parameters is
${\bf \Theta_R} = \{\tau_1, \ \tau_2, \ t_{\rm max},  \ \sigma_2 / \sigma_1\}$. 
The adopted  likelihood,  $\mathscr{L}$, used to compute the posterior probability distribution can be written as
\begin{equation}
\ln  \mathscr{L}  =-\sum_{n=1}^N\ln\left(\left(2 \pi \right)^{d/2} \prod_{j=1}^{d} \sigma_{n,j}\right)
-\frac{1}{2}\sum_{n=1}^N\sum_{j=1}^{d}\left(\frac{x_{n,j}-\mu_{n,j}}{\sigma_{n,j}} \right)^2,
\label{eq:likelihood}
\end{equation}
where $N$ is the number of stars in a Galactic region corresponding to $R$.
The quantities $x_{n,j}$ and $\sigma_{n,j}$ are respectively the measured value of $j^{\rm th}$ 
observable and its uncertainty for $n^{\rm th}$ star.   $\mu_{n,j}$ is  the model value of the $j^{th}$ observable for the $n^{th}$ star. As underlined in \citet{spitoni2020}, the curve predicted by the two-infall model in the plane [Mg/Fe] vs. [Fe/H] is 
multi-valued (see their Fig. 1). As a result, 
an observed data point in the
[Fe/H]-$[\alpha/{\rm Fe}]$ plane cannot be associated unambiguously to a point on the curve.
 To get through this problem they  associated  a data point  to the closest value on the curve  given a data point $x_{n,j}$,  defining the following  "distance data-model" function $D$,

\begin{equation}
D_{n,i} \equiv \sqrt{\sum_{j=1}^{d} \left( \frac{x_{n,j}-\mu_{n,j,i}}{\sigma_{n,j}} \right)^2}
\label{eq:distance1}
\end{equation} 
where $i$ runs over a set of discrete  values on the curve. 
Hence, the closest point on the curve  is $\mu_{n,j} = \mu_{n,j,i'}$ which fulfils the following relation:
\begin{equation}
S_n \equiv \min_i\left\{ D_{n,i}\right\}= \sqrt{\sum_{j=1}^{d} \left( \frac{x_{n,j}-\mu_{n,j,i'}}{\sigma_{n,j}} \right)^2}.
\label{eq:distance1}
\end{equation}

Here, we  present the priors on ${\bf \Theta_R} = \{\tau_1, \ \tau_2, \ t_{\rm max}, 
\ \sigma_2 / \sigma_1\}$. We use uniform priors for all parameters, which are independent of the Galactocentric distances (i.e priors are the same for all model runs at different $R$).
In the classical two-infall model \citep{chiappini1997, spitoni2020}, the first gas infall 
    is characterized by a short time-scale of accretion in the solar neighborhood, fixed at the value of $\tau_1$= 1 Gyr. More recently,  \citet[][ in order to reproduce the AMBRE thick disc]{grisoni2017}, and 
    \citet{spitoni2019} suggested a smaller value,  $\tau_1=0.1$ Gyr. 
    In the current study, we set a uniform prior
    on $\tau_1$ exploring the range $0 < \tau_{1} < 7  \mbox{ Gyr}$.
   The second infall time-scale,  $\tau_2$,   is 
    connected to a slower accretion episode. We set a uniform prior on  $\tau_2$ 
    exploring the range $0 < \tau_2 < 14  \mbox{ Gyr}$. For  the delay $t_{\rm max}$ we set a uniform prior 
    exploring the range $0 < t_{\rm max} < 14  \mbox{ Gyr}$.

\begin{figure}
\begin{centering}
\includegraphics[scale=0.37]{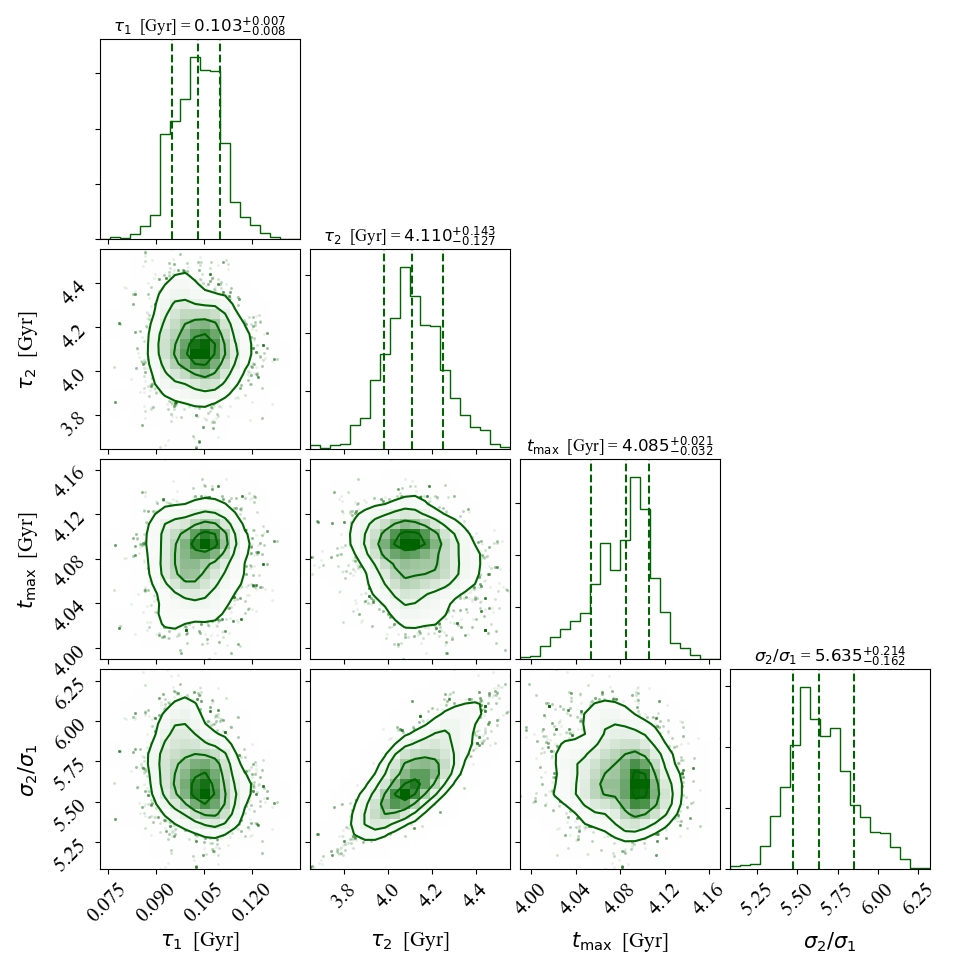}
\caption{Corner plot showing the posterior PDFs of the chemical evolution model parameters for the region at 8 kpc. For each parameter, the median and
the 16th and 84th percentiles of the posterior PDF are shown with dashed lines above the
marginalised PDF. The SFEs are fixed at values of 2 and 1 Gyr$^{-1}$ for the high- and low-$\alpha$ phases, respectively.}
\label{corner8}
\end{centering}
\end{figure}

Concerning    the present-day ratio between the total surface mass densities  $\sigma_2/\sigma_1$, we recall that 
    \citet{fuhr2017} derived  in the solar vicinity a local mass density 
    ratio between the thin and thick disc stars of 5.26, which becomes as low as 1.73 after correction for the 
    difference in the scale height. While studying APOGEE stars, \citet{mac2017} found that the relative 
    contribution of low- to high-$\alpha$ is 5.5. 
    In the solar annulus,  \citet{spitoni2020} found that the best models span the range between $3.2-4.3$.  
    In this work, studying different Galactic regions, we set  this prior in the range, $0.1 < \sigma_2/\sigma_1 < 50$.

The affine invariant MCMC ensemble sampler, "\textit{emcee}: the mcmc hammer" code\footnote{\href{https://emcee.readthedocs.io/en/stable/}{https://emcee.readthedocs.io/en/stable/};
\href{https://github.com/dfm/emcee}{https://github.com/dfm/emcee}}, proposed by  \citet{goodman2010,foreman} has been used to  sample the posterior probability distribution. 

\begin{figure}
\begin{centering}
\includegraphics[scale=0.38]{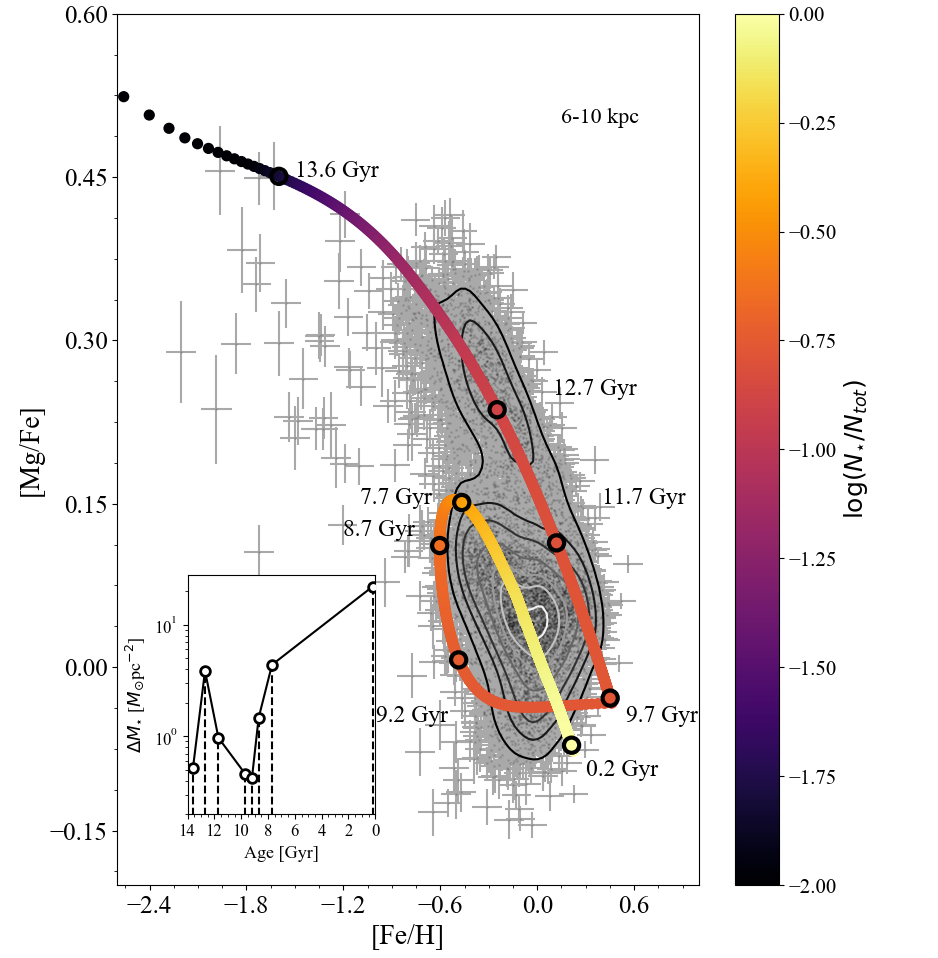}
\caption{Observed [Mg/Fe] vs. [Fe/H] abundance ratios from APOGEE DR16 \citep{Ahumada2019} (grey points with associated errors) in the  Galactocentric region  between 6 and 10 kpc compared with the best-fit chemical evolution model (thick curve) in that region. As in Fig. \ref{apogee_data}, the contour lines enclose fractions of 0.90, 0.75, 0.60, 0.45, 0.30, 0.20, 0.05 of the total number of observed stars.
The color coding represents the cumulative number of stars formed  during the Galactic evolution normalized to the total number $N_{tot}$.
%(the time-step is constant and fixed at the value of $\Delta \, t=4.6 \cdot 10^{-3}$ Gyr).
The open circles mark the model abundance ratios of stellar populations with different ages. 
In the inset we show the surface 
stellar mass density $\Delta M_{\star}$ formed in different age bins as a function of age, where the bin sizes are delimited by the vertical dashed lines and correspond to the same age values as indicated in the [Mg/Fe] vs [Fe/H] plot. 
}
\label{mg8}
\end{centering}
\end{figure}

\section{Results} \label{results}

Here,  we show predictions of chemical evolution models of the Galactic disc computed at different Galactocentric distances. In Section \ref{sec_out} we present and discuss our findings for chemical evolution models computed at 8 and 12 kpc, whereas the innermost region centered at 4 kpc is presented in Section \ref{sec_in}.
In Section \ref{sec_inside} we interpret our results in terms of the "inside-out" formation scenario and we show a comparison between our predictions and some of the most important observables of the Galactic disc. In Section  \ref{sec_MDF} the metallicity and the [Mg/Fe] distributions will be presented.
Finally, in Section  \ref{palla} we compare our findings with the  chemical evolution predictions of the recent study presented by \citet{palla2020}.

\subsection{Outer disc evolution: a tale of gas accretion and dilution}\label{sec_out}

First, in this section we present the results of the best-fit chemical evolution model at 8 kpc, which was obtained by fitting the abundance ratios [Mg/Fe] and [Fe/H] of stars in the APOGEE DR16 sample using our Bayesian technique based on the MCMC methods.

This model  assumes infall episodes with primordial chemical composition  for both high-$\alpha$ and  low-$\alpha$ sequences ($\mathcal{X}_{1}$ and $\mathcal{X}_{2}$ quantities in eq. \ref{infall}) and different SFE: $\nu_1=$ 2 Gyr$^{-1}$ and  $\nu_2=$ 1 Gyr$^{-1}$ \citep{chiappini1997,grisoni2018,spitoni2020, palla2020}. 
In Fig. \ref{corner8} we  show the  posterior probability density function (PDF) of the chemical evolution model parameters, 
 ${\bf \Theta_{\odot}} = \{\tau_1, \ \tau_2, \ t_{\rm max}, \ \sigma_2 / \sigma_1\}$, for our model at a Galactocentric distance of 8 kpc. 
 We  find  a significant delay in the start of the second gas infall $t_{{\rm max}}=4.085^{+0.021}_{-0.032}$ Gyr, confirming the previous results of  \citet{spitoni2019,spitoni2020}.
The best model  predicts a value of  5.635$^{+0.214}_{-0.162}$ for the $\sigma_2/\sigma_1$  ratio, in accordance  with the findings of  \citet{mac2017,fuhr2017,spitoni2020}.
Predicted infall time-scales $\tau_1=0.103^{+0.007}_{-0.007}$ Gyr and $\tau_2=4.110^{+0.143}_{-0.127}$ Gyr are shorter than the ones of  \citet{spitoni2020}. We recall that in that work the model was compared  with the APOKASC  sample  by \citet{victor2018}, which consisted of 1180 red giants in a narrower region in the solar vicinity (2 kpc-wide). The APOKASC sample has measured solar-like oscillations, allowing asteroseismic determination of stellar ages. In \citet{spitoni2020}, the determined stellar ages were used as additional constraint.
Recall that here we adopt different selection criteria  for  stars following \citet{weinberg2019} and \citet{vincenzo2020}.

In fact,  while the APOKASC sample shows absence of stars with  [$\alpha$/Fe]<-0.05 dex (see Fig. 1 in \citealt{spitoni2020}), the sample adopted here has a significant fraction of low-$\alpha$ stars with [Mg/Fe]<-0.05 dex. Hence, the differences in the best-fit values for the infall time-scales can be attributed to the differences in the data used in the two studies.

\begin{table}[]
\begin{center}
\caption{Observed solar chemical abundances compared with   model predictions.}
\label{tab_sol}
\begin{tabular}{c|cc}

  \hline
  \hline
%\noalign{\smallskip}

 Abundance &{\it Observations} &  {\it Model} \\
 $\log$($X$/H)+12 & \citet{asplund2005}&$R$=8 kpc\\
 
 & [dex]&[dex] \\
%\noalign{\smallskip}
%\noalign{\smallskip}
\hline
\\
%\noalign{\smallskip}
Fe &7.45$\pm$0.05&7.40\\
Mg& 7.53$\pm$0.09 &7.51\\
Si& 7.51$\pm$0.04 &7.49\\
%\noalign{\smallskip}
 \hline
%\noalign{\smallskip}
\end{tabular}
\end{center}
\end{table}

In Fig. \ref{mg8}  we  compare the best-fit model computed at 8 kpc with the [Mg/Fe] vs. [Fe/H] abundance ratios.
The color coding with the cumulative number of stars formed  during the Galactic evolution  shows that the bulk of the stars are formed during the low-$\alpha$ sequence.
Moreover, the surface stellar mass density  $\Delta M_{\star}$ formed in different age bins as a function of age (see the inset plot of Fig. \ref{mg8}), clearly shows the existence of a low-$\alpha$ and high-$\alpha$ bimodality.

For the sake of clarity, in response to a question posed in \citet{lian2020b} we underline that this bimodality was already present in the best fit model of  \citet{spitoni2019} who did not show the variation of the stellar mass content along with the chemical evolution tracks in their figures.
\begin{figure}
\begin{centering}
\includegraphics[scale=0.37]{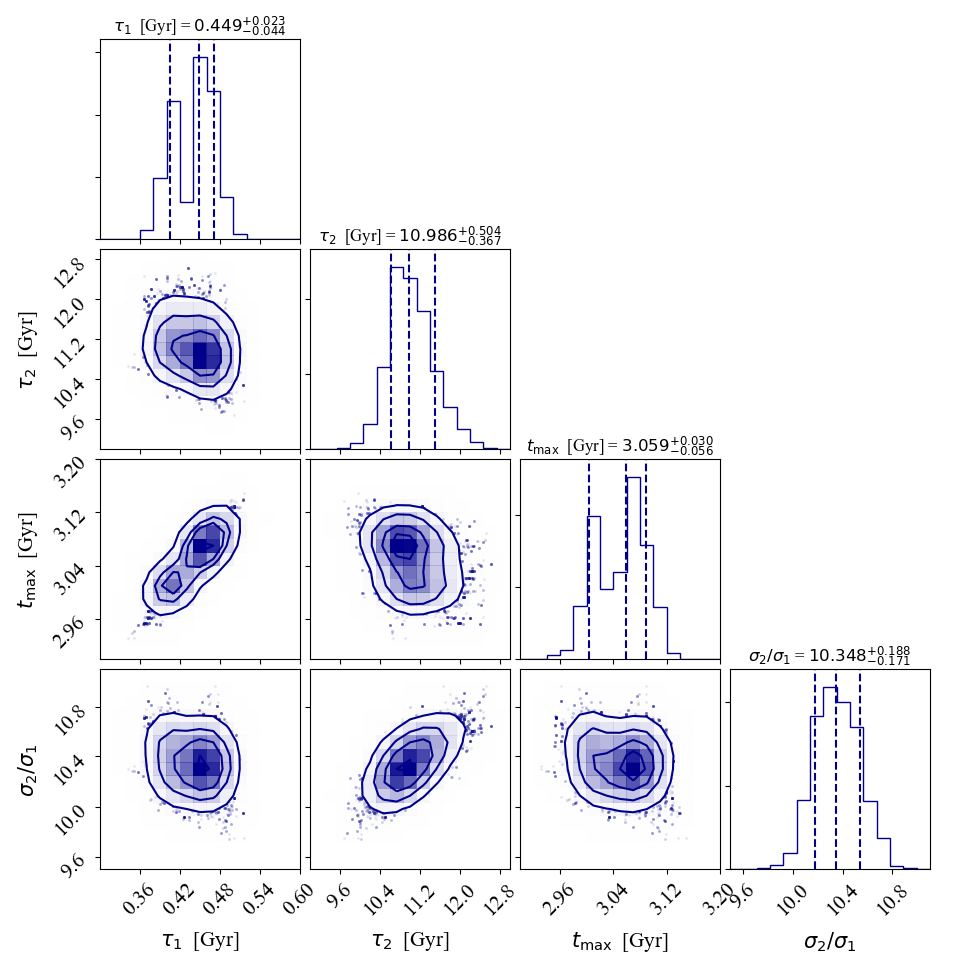}
\caption{
Same as Fig. \ref{corner8}, but for the annular region enclosed between 10 and 14 kpc. The SFEs are fixed at values of 2 and 0.5 Gyr$^{-1}$ for the high-$\alpha$ and low-$\alpha$ phases, respectively.}
\label{corner12}
\end{centering}
\end{figure}

\begin{figure}
\begin{centering}
\includegraphics[scale=0.38]{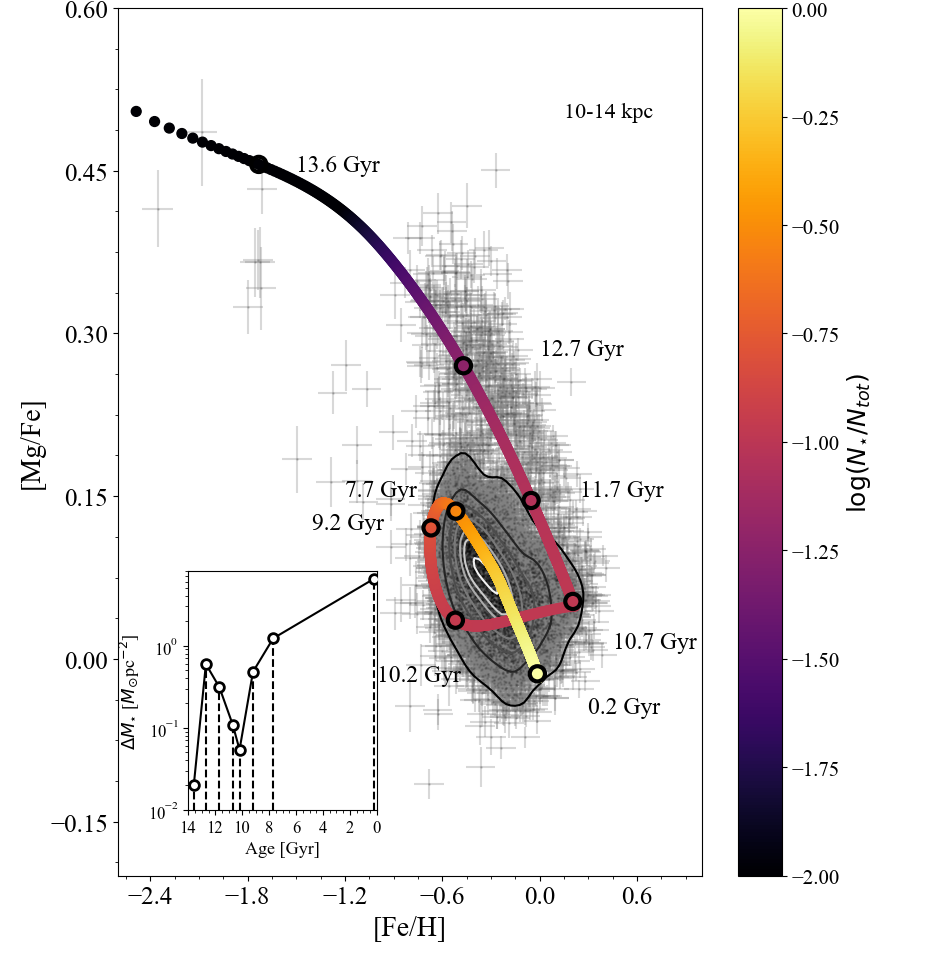}
\caption{Same as Fig. \ref{mg8}, but for the observed abundance [Mg/Fe] vs. [Fe/H] ratio for stars  with Galactocentric distances between 10 kpc and 14 kpc and the corresponding best-fit chemical evolution model.}
\label{mgfe12}
\end{centering}
\end{figure}

As  illustrated in \citet{spitoni2020,spitoni2019}, the gas dilution originated by a strong second gas infall is a
key process to explain  APOKASC abundance ratios with our model. 
In particular, the  second accretion  event of  pristine gas decreases the metallicity of the stellar populations born immediately after  keeping  a roughly constant [Mg/Fe] ratio since the accretion involves H and He but basically no metals.
When star formation resumes,  Type II SNe produce a steep bump in  [Mg/Fe], which subsequently decreases at higher metallicities due to iron  from Type Ia SNe \citep{matteucci2009, bonaparte2013}. This sequence produces a loop  feature in the chemical evolution track of [Mg/Fe] vs. [Fe/H]. 
  We  notice from the inset plot of Fig. \ref{mg8} that a negligible  mass fraction of low-$\alpha$ stars  is formed during the  dilution phase of the loop, with ages in the range between 9.7 and 9.2 Gyr ($\Delta M_{\star}$= 0.42 M$_{\odot}$ pc$^{-2}$, corresponding to $\sim$ 1.49\%  of
of the stellar mass formed during the second infall episode).  
In the ascending part of the loop with ages  between 9.2 and 8.7 Gyr,  the formed stellar mass is    $\Delta M_{\star}$= 1.46 M$_{\odot}$ pc$^{-2}$  ($\sim$ 5.18\%  of the predicted low-$\alpha$ sequence stellar mass).
Finally, almost   the totality of stellar mass is produced in the age interval  between 8.7 Gyr and 0.2 Gyr, namely  $\Delta M_{\star}$=26.31  M$_{\odot}$ pc$^{-2}$, corresponding to $\sim$ 93.3\%  of the entire stellar mass produced during the second gas infall.

The ascendant part  of the loop   seems to pass through a region with not so many stars. We underline that the  availability of additional observables in the MCMC analysis, as for instance precise asteroseismic ages, can potentially alleviate this apparent tension, leading to a smaller loop in the [$\alpha$/Fe] vs. [Fe/H] space, as shown in \citet{spitoni2020}. 

We  stress that, as shown in   \citet{spitoni2019},  it is likely that the loop feature is  hidden inside the observational errors.
Using a  "synthetic" model (at each Galactic time 
a random error  was assigned to the ages and metallicities of the stellar populations),   \citet{spitoni2019} were capable to  reproduce the data spread  in the low-$\alpha$ sequence  of the APOKASC sample \citep{victor2018} in the [$\alpha$/Fe] vs. [Fe/H] abundance ratio space. 

\begin{figure}
\begin{centering}
\includegraphics[scale=0.37]{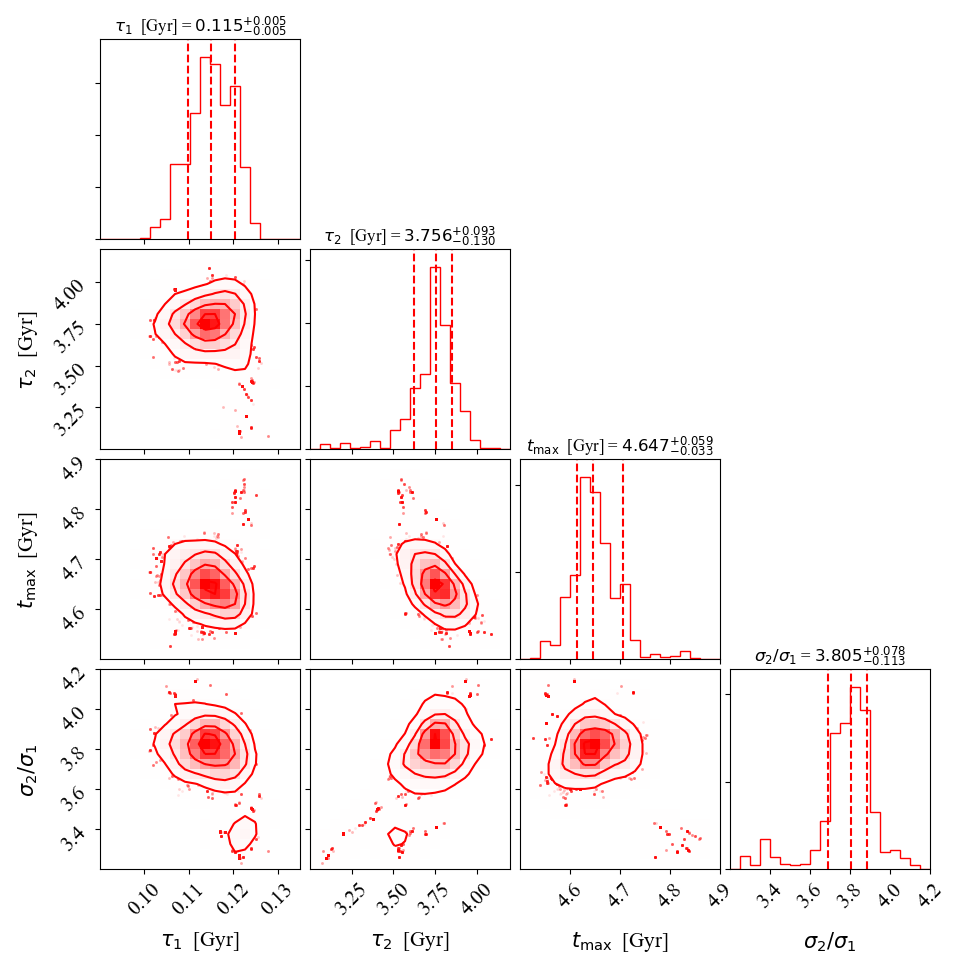}
\caption{
Same as Fig. \ref{corner8}, but for the annular region enclosed between 2 and 6 kpc. For the second gas accretion episode,  an enriched infall with  [Fe/H]= -0.5 dex is considered. The SFEs are fixed at values of 3 and 1.5 Gyr$^{-1}$ for the high-$\alpha$ and low-$\alpha$ phases, respectively.}
\label{corner4_SFE3}
\end{centering}
\end{figure}

\begin{figure}
\begin{centering}
\includegraphics[scale=0.38]{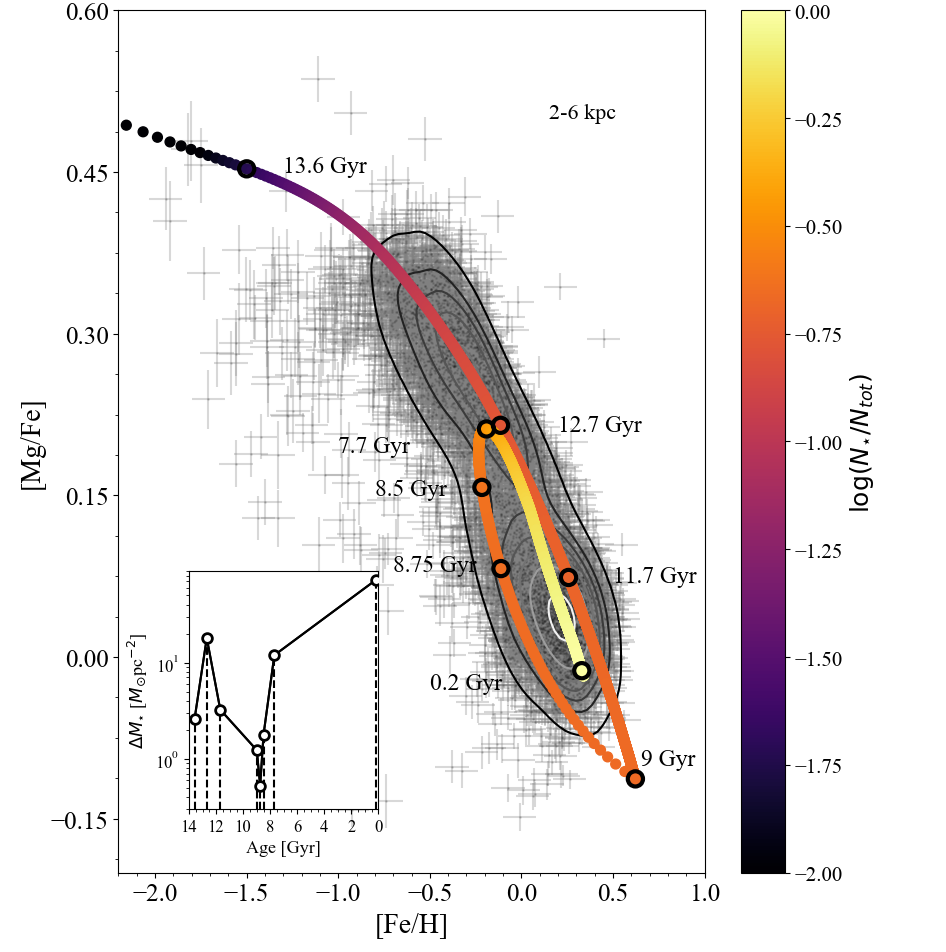}
\caption{Same as Fig. \ref{mg8}, but for stars of APOGEE DR16 \citep{Ahumada2019}  sample with Galactocentric distances between 2 and 6 kpc and  the corresponding best fit chemical evolution model. Here, we considered a pre-enriched second gas infall with metallicity, [Fe/H]= -0.5 dex. The SFEs are fixed at values of 3 and 1.5 Gyr$^{-1}$ for the high- and low-$\alpha$ phases, respectively.}
\label{mg4_SFE3}
\end{centering}
\end{figure}

In Table \ref{tab_sol}, we compare the solar abundances of Fe, Mg, and Si predicted by our best fit model in the solar neighborhood  with \citet{asplund2005} photospheric values. In the model, solar abundances are determined from the composition of the ISM at the time of the formation of the Sun (after 9.5 Gyr from the Big Bang). It is evident that our model is able to  reproduce well  the solar abundance ratios for these elements.

In Figs.  \ref{corner12}  and  \ref{mgfe12} we show the results for the external  region centered at 12 kpc.
For this model we  also assume primordial infall for both the high- and  low-$\alpha$ sequences  and different SFEs for the high- and low-$\alpha$ phases, $\nu_1=$ 2 Gyr$^{-1}$ and  $\nu_1=$ 0.5 Gyr$^{-1}$, respectively.  As we stated in  Section \ref{presc}, a lower SFE in outer Galactic regions has been assumed by several chemical evolution models  \citep{colavitti2008, spitoni2015, grisoni2018,palla2020} in order to  reproduce abundance gradients (see Section \ref{sec_inside}).

Comparing  the best fit model parameters for the Galactic region centered  at 8 kpc with those at 12 kpc, we note that longer time-scales of gas accretion $\tau_1(R)$ and  $\tau_2(R)$  are associated to more external region. 
 Hence, the  `inside-out' formation scenario invoked by  the classical two-infall model of \citet{chiappini2001}  in our model  is a natural consequence of the fit to the APOGEE DR16 abundance ratio, as we shall discuss it thoroughly in Section \ref{sec_inside}.
 \subsection{Inner disc evolution: enriched gas infall and  starbursts}\label{sec_in}

Concerning the evolution of the inner disc, we consider the presence of an enriched gas infall for the low-$\alpha$ phase ($\mathcal{X}_{2}$ quantity in eq. \ref{infall}).
In \citet{palla2020}, it was already pointed out that a  primordial infall for the  inner  thin disc cannot explain the observed behaviour of the  [$\alpha$/Fe] vs. [Fe/H] in the APOGEE data.

Different physical reasons may be associated with this enriched gas infall event.
For instance,  \citet{palla2020}  discussed that an enriched infall could partly be due to  gas lost from the formation of the thick disc, Galactic halo or the Galactic bar which then gets mixed with a larger amount of infalling primordial gas as proposed by \citet{gilmore1986}.
In \citet{Khoperskov2020},  the infalling gas during inner thin disc formation phase is not primordial because the gaseous halo has been significantly polluted during the high-$\alpha$ disc formation, providing a tight connection between chemical abundance patterns in the high-$\alpha$ and low-$\alpha$ discs.
Alternatively,  as already mentioned in the Introduction,  \citet{VINTERGATANIII2020}  proposed that   the Galaxy disc is fueled by two distinct gas flows and that the one responsible for the formation of the inner $\alpha$ sequence is enriched by outflows from massive galaxies (with $\sim$ 1/3 the Milky Way mass at the time of the interaction).

Following  the best model prescriptions of \citet{palla2020}  for the inner thin disc, we impose that  for the second gas infall (low-$\alpha$) a chemical enrichment obtained from the model of the high-$\alpha$ disc phase  corresponding to  [Fe/H]=-0.5 dex. 
In Figs. \ref{corner4_SFE3} and \ref{mg4_SFE3} we present the chemical evolution model predictions  for the region centered at 4 kpc assuming, for the high-$\alpha$ sequence, a SFE fixed at the value of $\nu_1=$ 3 Gyr$^{-1}$. This higher value compared to external regions could be motivated by starburst episodes, as suggested by \citet{VINTERGATANI2020, VINTERGATANII2020,VINTERGATANIII2020}.
The time-scale $\tau_2$ of the second gas accretion episode and the ratio between low-$\alpha$ and high-$\alpha$ surface mass density $\sigma_2/ \sigma_1$ are smaller compared to the external parts, in perfect agreement with the inside-out formation scenario (see Section \ref{sec_inside} for a detailed discussion).

 We also explore the possibility of a SFE  fixed at the value of 2 Gyr$^{-1}$ which as the same as in the outer regions. Although  we found a similar   [Mg/Fe] vs. [Fe/H] relation, an extremely short (and maybe unrealistic)   infall time-scale of $\tau_1=$ 0.007 Gyr is required. We underline that such  time-scale is  too small even compared to   the Galactic bulge accretion time-scale \citep{matteucci2019,matteucci2020}.  Hence,  a  higher SFE   ($\nu_1 \sim$ 3 Gyr$^{-1}$) in the high-$\alpha$ disc phase provides  more reasonable results   in the two-infall framework.

 \begin{table*}[]
\begin{center}
\tiny
\caption{
In the upper part of the Table we show
the chemical composition of the two gas infall episodes ($\mathcal{X}_{1}$ and $\mathcal{X}_{2}$) and the star formation efficiencies ($\nu_1$ and  $\nu_2$) for the high-$\alpha$ and low-$\alpha$ sequences assumed in our model at various Galactocentric distances. 
In the lower part of the Table we show the  
 accretion time-scales ($\tau_1$ and  $\tau_2$), 
the present-day total surface mass density ratio ($\sigma_{2}$/ $\sigma_{1}$) and delay $t_{{\rm max}}$
computed for our best fit models at 4 kpc, 8 kpc and 12 kpc (see text for model details). In the last column on the right, we also provide the ranges admitted by our study computed from our best-model estimates  at all Galactocentric radii. 
}
\label{tab_mcmc}
\begin{tabular}{c|ccc|c}
\hline
  \hline
%\noalign{\smallskip}
% &  \multicolumn{3}{c}{\it Models}  & \\
 &  &{\it Models} &&\\
 &  &&&\\

  & 4 kpc&8 kpc &12 kpc&\\
  
\hline

$\mathcal{X}_{1}$ &Primordial&Primordial&Primordial\\
&&&\\
$\mathcal{X}_{2}$ &[Fe/H]=-0.5 dex&Primordial&Primordial\\
&&&\\
  $\nu_1$  [Gyr$^{-1}$]  &3.0&2.0&2.0  \\
  &&&\\
   $\nu_2$  [Gyr$^{-1}$]   &  1.5&1.0&0.5\\

\hline

    & &{\it MCMC Results}&&Range\\
  \hline

 $\tau_{1}$ [Gyr]&0.115$^{+0.005}_{-0.005}$&0.103$^{+0.007}_{-0.006}$& 0.449$^{+0.023}_{-0.044}$&$0.097-0.472$\\

&&&\\
$\tau_{2}$ [Gyr]&3.756$^{+0.093}_{-0.130}$&4.110$^{+0.145}_{-0.127}$&  10.986$^{+0.504}_{-0.367}$& $3.626-11.490$\\

&&&\\
 $\sigma_{2}$/ $\sigma_{1}$&3.805$^{+0.078}_{-0.113}$&5.635$^{+0.214}_{-0.162}$&  10.348$^{+0.188}_{-0.171}$&$3.692-10.536$\\
 &&&\\
 $t_{\rm max}$ [Gyr]&4.647$^{+0.059}_{-0.033}$&4.085$^{+0.021}_{-0.032}$& 3.059$^{+0.030}_{-0.056}$&$3.003-4.706$\\

%\noalign{\smallskip}
 \hline
%\noalign{\smallskip}
\end{tabular}
\end{center}
\end{table*}

\subsection{Inside-out formation scenario and global properties of the Galactic disc}\label{sec_inside}
In Table \ref{tab_mcmc} we summarize the best fit model parameters at different Galactocentric distances:  the delay $t_{{\rm max}}$, surface density ratio $\sigma_{2}$/$\sigma_{1}$, infall time-scales $\tau_1$ and $\tau_2$ values.
The  presence of a significant  delay between the two infall episodes is a robust result, confirming the previous results of the works focusing on the solar neighborhood \citep{spitoni2019, spitoni2020}. 

\begin{figure}
\begin{centering}
\includegraphics[scale=0.4]{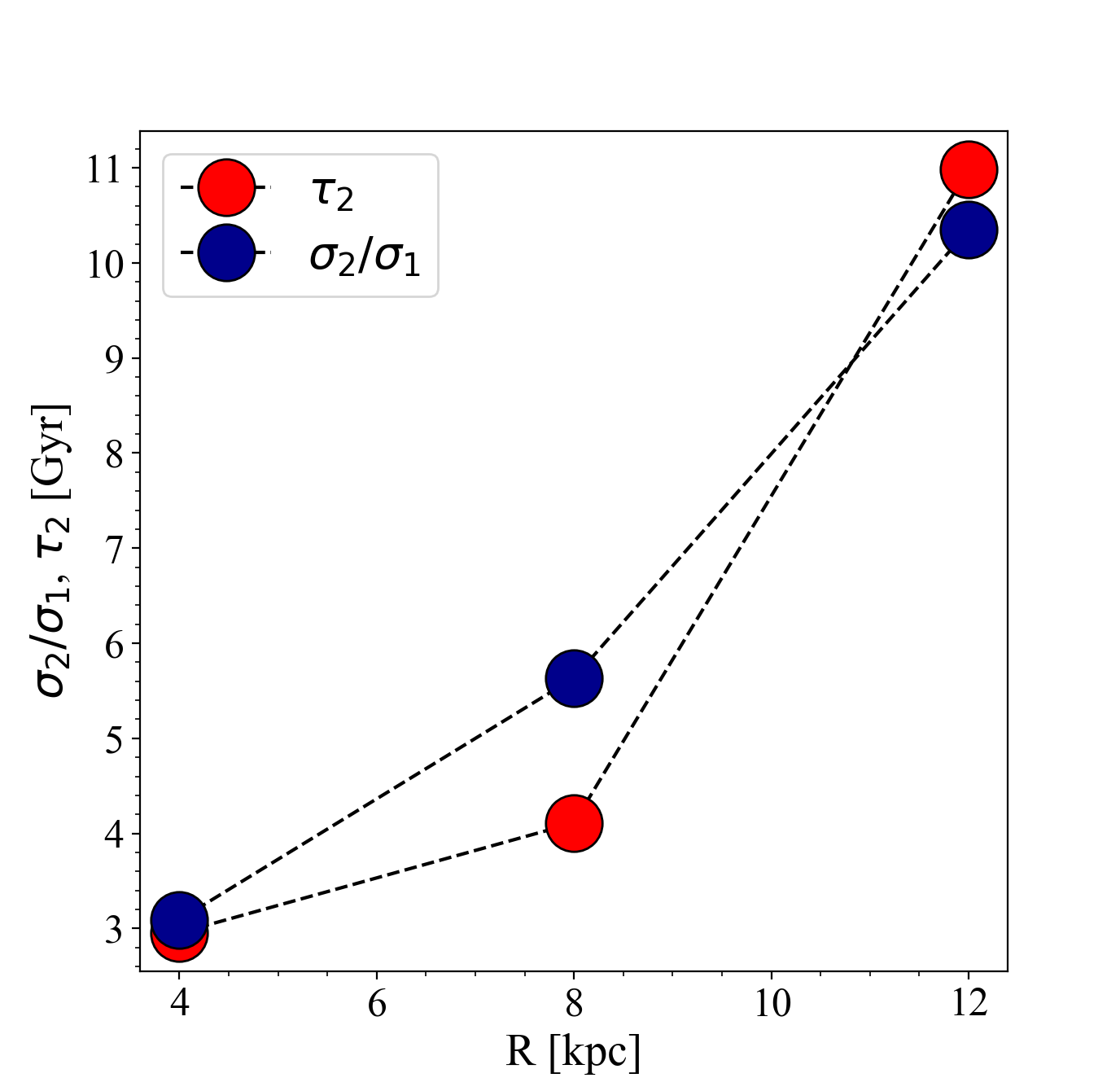}
\caption{ Time-scale $\tau_2$ for the gas accretion in the  low-$\alpha$ phase  and the ratio $\sigma_2/\sigma_1$ between the  low-$\alpha$ and  high-$\alpha$ total surface mass densities  as a function of the Galactocentric distance are drawn with red and blue points, respectively. }
\label{inside}
\end{centering}
\end{figure}

In Fig. \ref{inside} we show that our model predictions are in favour of the 'inside-out' formation scenario: inner Galactic regions are assembled faster compared to the external one \citep{matteucci1989,chiappini2001,schoenrich2017,frankel2019}.
In Fig. \ref{apogee_data} data show that in the outer regions the locus of the low-$\alpha$ sequence shifts towards lower metallicity and  in Fig. \ref{inside} we find that external Galactic regions are formed on longer accretion  time-scales $\tau_2$, hence the chemical enrichment is weaker and less efficient than the inner Galactic regions, leading to a lower metallicity. We recall  that we imposed also a radial variation of the SFE.

Moreover, in Fig. \ref{apogee_data}  the radial bin between 10 and 14 kpc   has fewer high-$\alpha$  stars  and we associate the  more prominent low-$\alpha$ sequence   to a  larger surface density ratio $\sigma_{2}$/$\sigma_{1}$  compared to the innermost regions in agreement with \citet{palla2020}. 
An important result of this study is that extending  the predictions for $\sigma_{2}$/$\sigma_{1}$ to the whole Galactic disc, we predict a clear and neat trend: the ratio increases with the Galactocentric distance (Fig. \ref{inside}).

\begin{figure}
\begin{centering}
\includegraphics[scale=0.4]{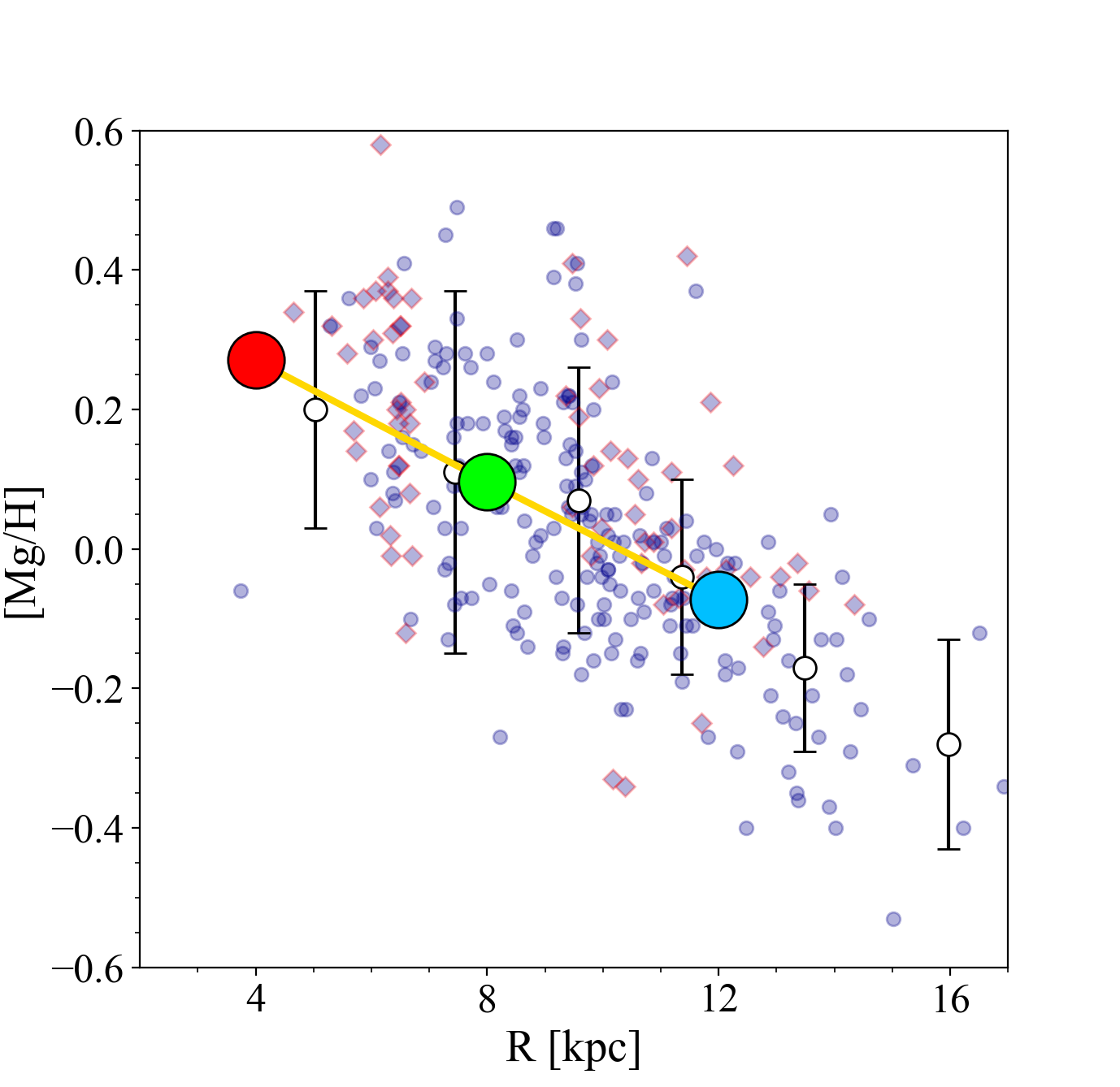}
\caption{Observed and predicted radial [Mg/H]  present-day abundance  gradient. The prediction of  the best-fit  models are indicated with big filled circles, connected with a yellow line.
The  observational data  are the Cepheid observations  from \citet{luck2011} (blue  circles) and 
\citet{genovali2015} (blue diamonds with red edges). With the empty circles  we report the average values and
associated errors of the full data  sample.
}
\label{grad}
\end{centering}
\end{figure}
\begin{figure}
\begin{centering}
\includegraphics[scale=0.4]{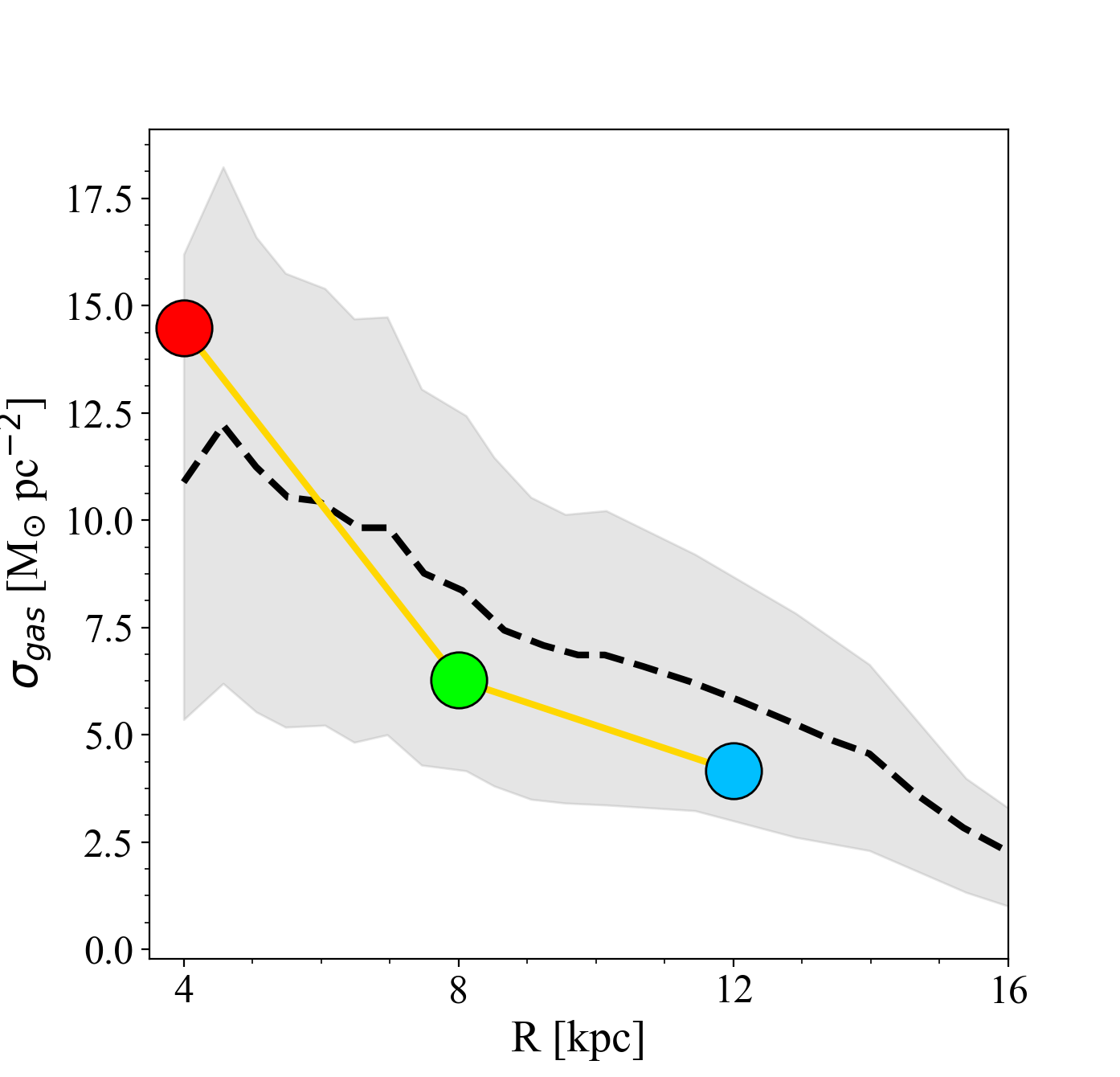}
\caption{Observed and predicted radial gas surface density gradient. The black dashed curve is the
average  between the \citet{dame1993} and \citet{Nakanishi2003,Nakanishi2006}
data sets. The grey shaded region represents the typical uncertainty at each radius,
for which we adopt either 50\% of the average (see \citealt{Nakanishi2006})
or half the difference between the minimum and maximum values in each
radial bin (if larger). The big filled  circles show the model predictions.}
\label{gas}
\end{centering}
\end{figure}

\begin{figure}
\begin{centering}\includegraphics[scale=0.4]{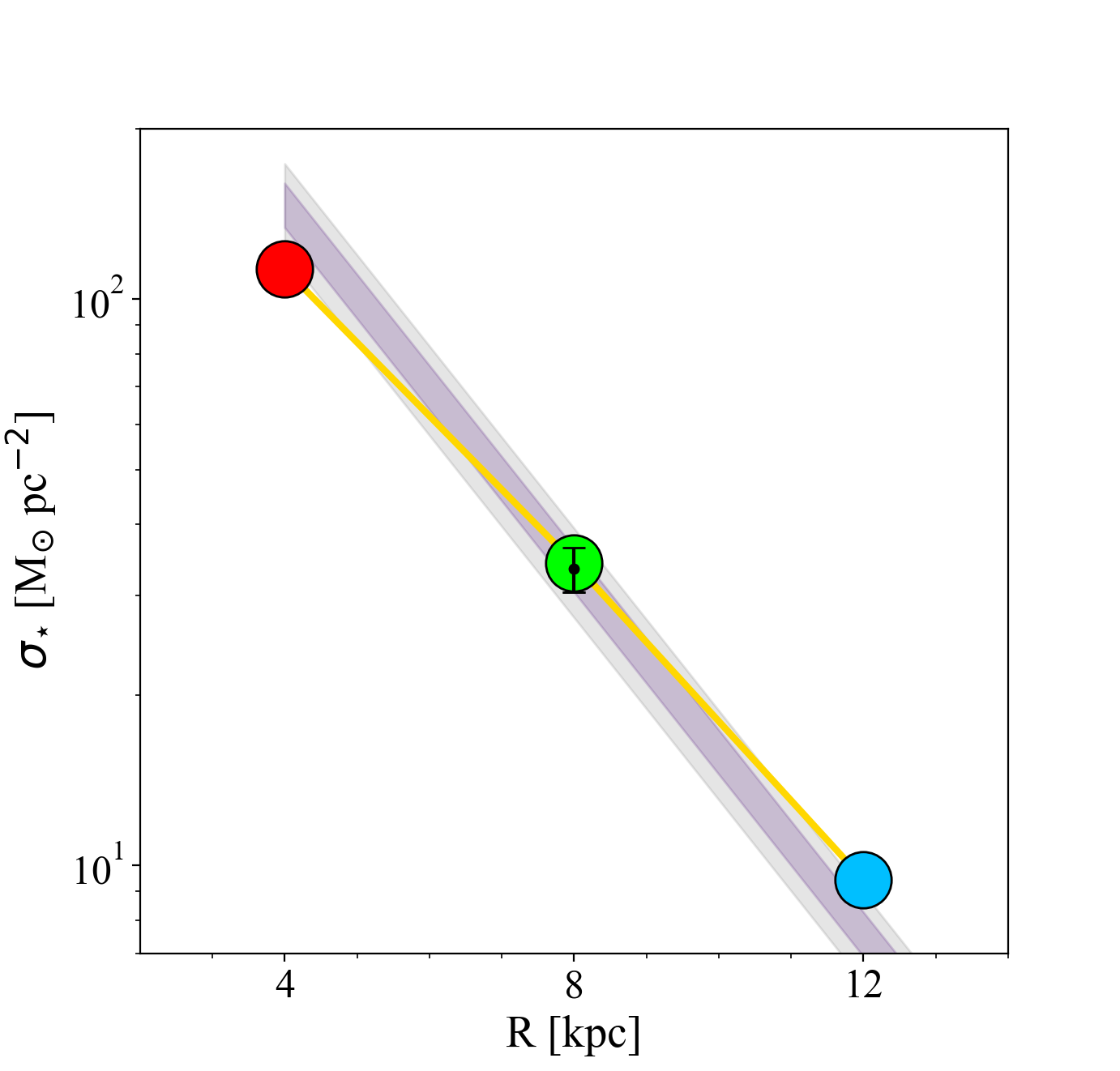}
\caption{
Radial stellar surface stellar density profile of our model  (big circles connected by a yellow line). 
The observed  local stellar density is  $33.4 \pm 3$ M$_{\odot}$ pc$^{-2}$
\citep[small black dot and associated error bar]{mckee2015}. In our model we assume that the stellar profile decreases exponentially outward with a characteristic scale-length of 2.7 kpc \citep{kubryk2015}. The blue  and grey shaded areas indicate  zones within  1$\sigma$ and 2$\sigma$, respectively.     }
\label{stars}
\end{centering}
\end{figure}
\begin{figure}
\begin{centering}
\includegraphics[scale=0.26]{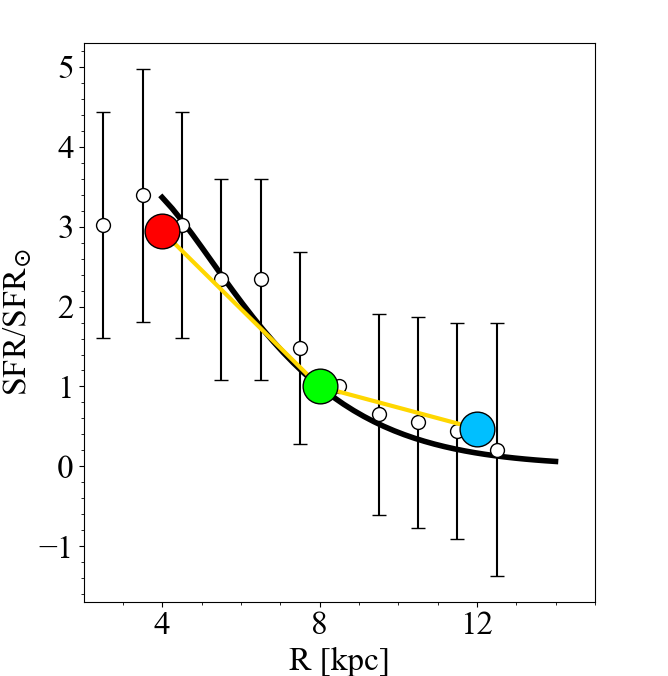}
\includegraphics[scale=0.26]{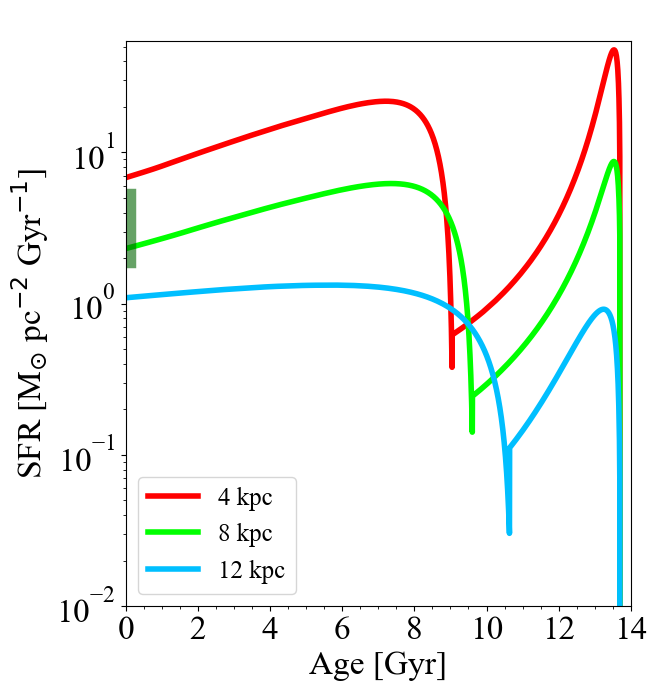}
\caption{ {\it Left panel}:
Observed and predicted radial SFR density gradient relative to the solar neighbourhood. 
The model results are the big solid circles connected by a solid yellow line. The black line is the analytical form  suggested by \citet{green2014} for the Milky Way SFR profile. The open circles with error bars are observational data from \citet{rana1991}.  
      {\it Right panel}: SFR evolution predicted by our best fit models computed at 4, 8 and 12 kpc. The dark green shaded area indicates the present-day measured range in the solar annulus by \citet{prantzos2018}.}
\label{SFR}
\end{centering}
\end{figure}

\begin{figure}
\begin{centering}
\includegraphics[scale=0.3]{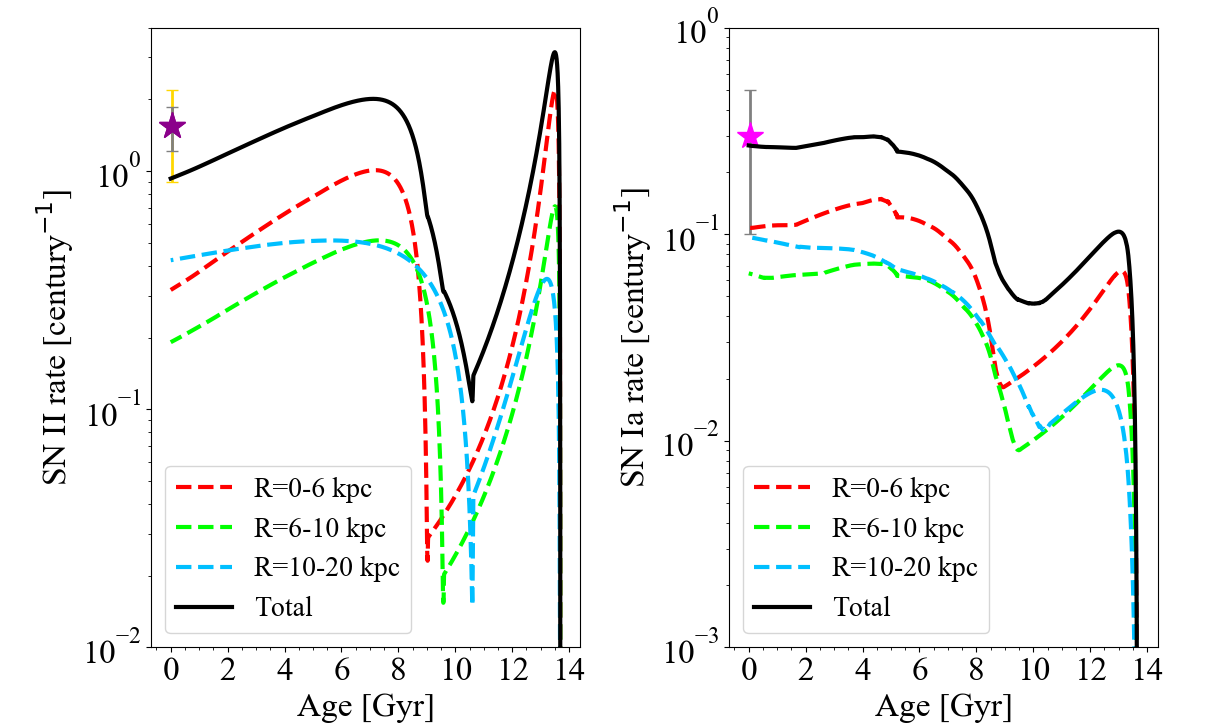}
\includegraphics[scale=0.25]{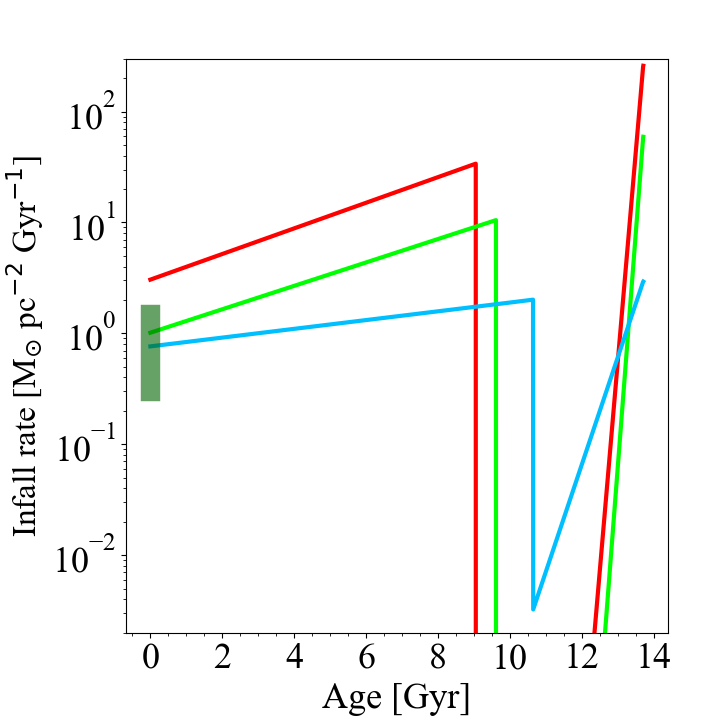}
\caption{ {\it Upper panels}:
Observed and predicted Type II SN rates (left) and Type Ia SN rates (right) as a function of the Galactic age.
The SN rates predicted for the whole disc are reported with the black solid lines, and they represent the sum of the contributions from different Galactic regions  indicated with colored  dashed lines.   The  observed present-day Type II SN rate of \citet[][left panel]{li2010} for the whole Galaxy  is reported with the solid star (1$\sigma$ and 2$\sigma$ errors are indicated with   grey and yellow bars, respectively) whereas the solid star in the right panel stands for  Type Ia SN rate of \citet{cappellaro1997} with the associated  1$\sigma$ error bar.     {\it Lower panel}: Infall  rate evolution predicted by our best fit models computed at 4 (red line), 8 (green line) and 12 kpc (light-blue line). The dark green shaded area indicates the present-day  values in the solar annulus  suggested by \citet{matteucci2012}.}
\label{SN}
\end{centering}
\end{figure}

\begin{figure}
\begin{centering}
\includegraphics[scale=0.345]{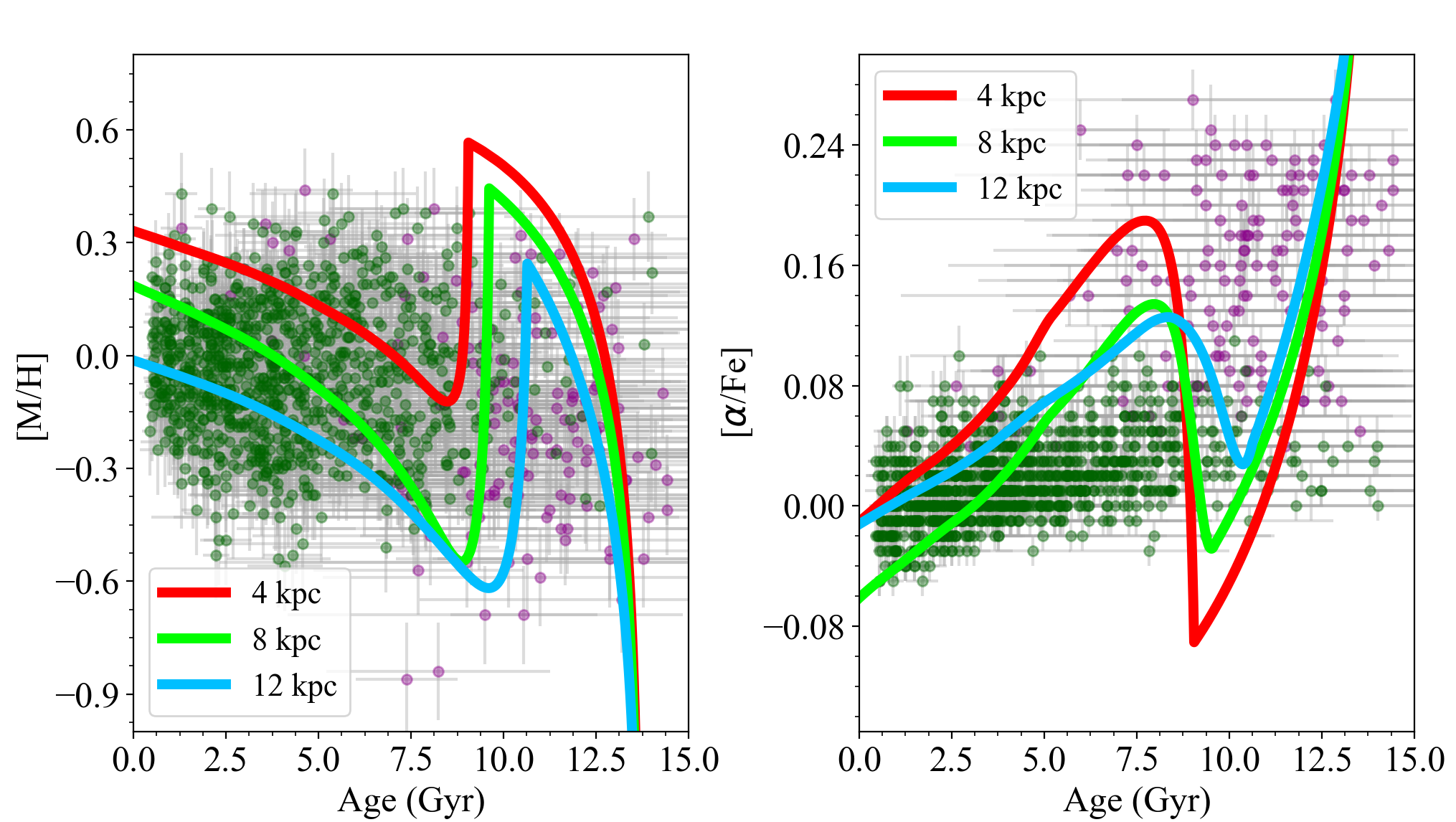}
\caption{Evolution of the  [M/H] (left panel) and [$\alpha$/Fe] (right panel)
abundance ratios of our best-fit models computed at 4, 8 and 12 kpc compared with the abundances observed in a stellar sample 
in the solar annulus  \citep{victor2018}. Magenta  points depict the high-$\alpha$ population, whereas green points indicate the low-$\alpha$ one. As in \citet{spitoni2019, spitoni2020}, we have not taken into account young $\alpha$-rich stars.}
\label{alpha_age}
\end{centering}
\end{figure}
 The Galactic 'inside-out' formation is well motivated by the dissipative collapse scenario \citep{larson1976,cole2000} and it has been  a widely adopted assumption coupled with a variable SFE and the radial gas flows in order to reproduce the observed abundance gradients \citep{spitoni2015, grisoni2018, palla2020}.

We have shown that  the two-infall model  of \citet{spitoni2019,spitoni2020} can be   extended to the whole disc admitting  a more   complex nature of the Galactic disc evolution instead of a simple sequential scenario, with  the coexistence of different physical processes and different  gas infall enrichment as a function of the Galactocentric distance. Moreover,  with a quantitative estimation of the model free parameters  using a Bayesian approach,  we confirm the results of the chemical evolution model proposed by \citet{palla2020}, where an enriched gas infall has been invoked to reproduce the chemical evolution of the inner Galactic disc (see Section \ref{palla} for a detailed comparison with \citealt{palla2020} predictions).
The proposed multi-zone chemical evolution model based on the MCMC methods to fit the abundance distributions of  [Mg/Fe] vs. [Fe/H] ratios from the  APOGEE DR16 sample is also able to reproduce the other important observables of the Galactic disc.

 In  Fig. \ref{grad} we compare the observed abundance gradient for magnesium of \citet{luck2011} and \citet{genovali2015} with  the one predicted by our best model computed  at the present-day.
To be consistent with these  data sets, the model abundance
ratios are referred to the solar value of \citet{grevesse1996}.
It is clear that our model prediction well reproduces the
observed  abundance gradient.

It is clear from Fig. \ref{gas} that also the observed surface gas density  profiles are  well reproduced by our model.
Concerning the present-day surface stellar mass density (see Fig. \ref{stars}), the predicted value at 8 kpc is 34.2 M$_{\odot}$ pc$^{-2}$ in very good agreement with the observed
local value of   $33.4 \pm 3$ M$_{\odot}$  \citep{mckee2015}. Assuming that the stellar surface density decreases exponentially outwards with a characteristic length-scale of  $R_{\star}$=2.7 kpc \citep{kubryk2015}, our model reproduces reasonably well this profile as shown in Fig. \ref{stars}.   
 
 \begin{figure}
\begin{centering}
\includegraphics[scale=0.36]{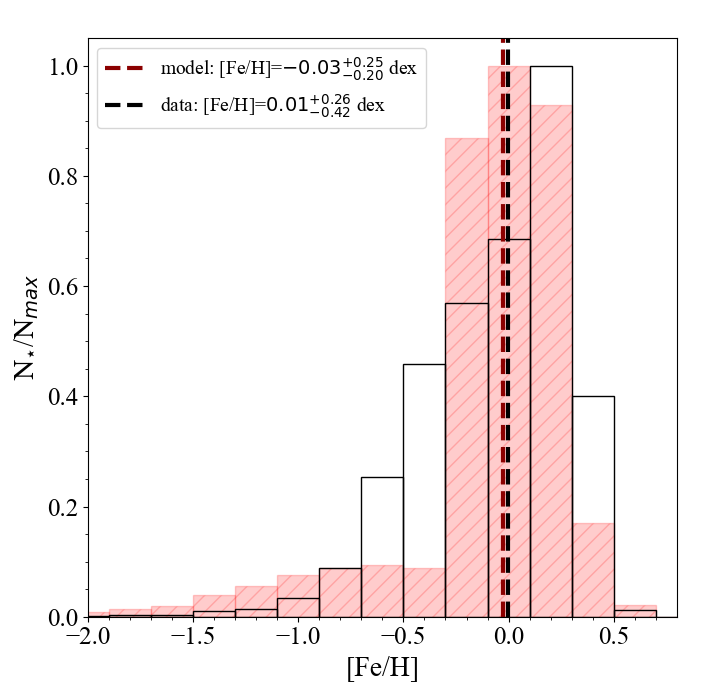}
\includegraphics[scale=0.36]{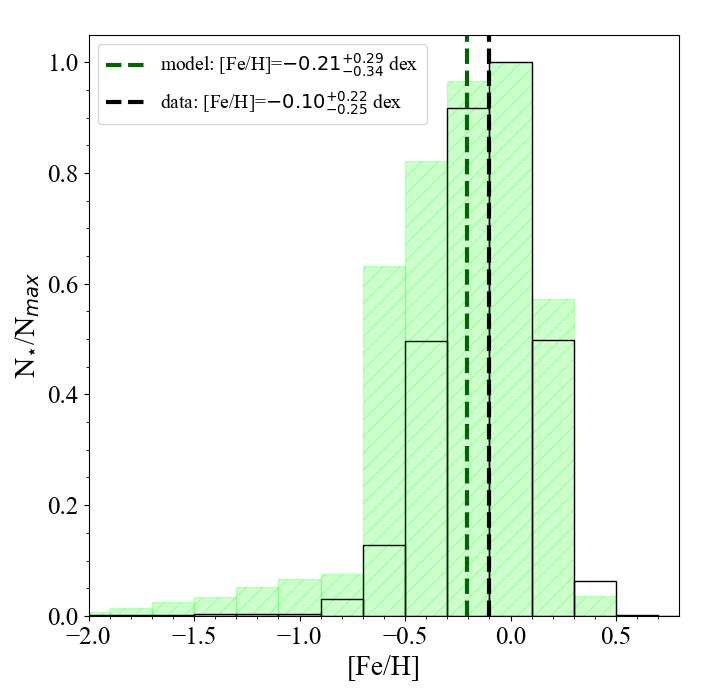}
\includegraphics[scale=0.36]{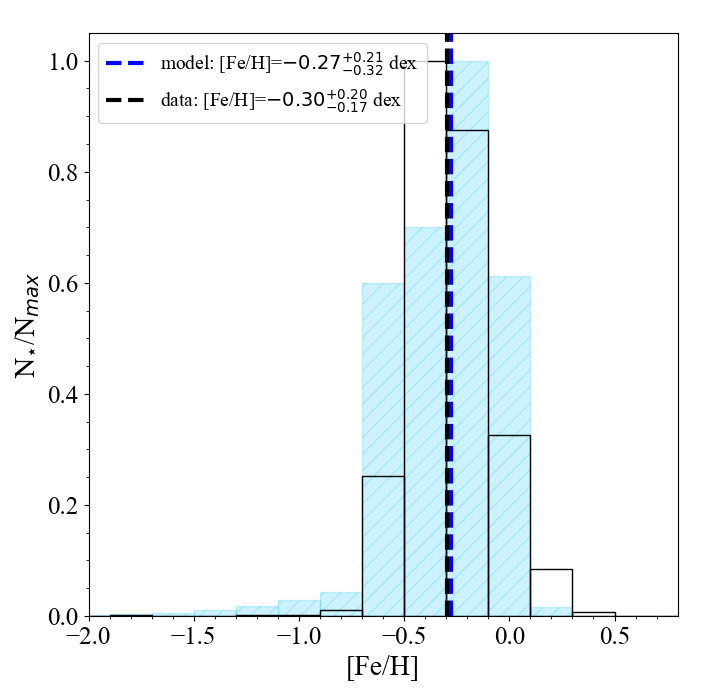}
\caption{Metallicity distributions predicted by the best-fit models (colored histograms) computed at 4 kpc (upper panel), 8 kpc  (middle panel), and 12 kpc (lower panel). The observed APOGEE DR16 distributions are shown by the black empty histograms. The  vertical lines indicate the median values of each distribution. In each plot, the distributions are normalised to the corresponding maximum number of stars, $N_{max}$.}
\label{MDF}
\end{centering}
\end{figure}

\begin{figure*}
\begin{centering}
\includegraphics[scale=0.365]{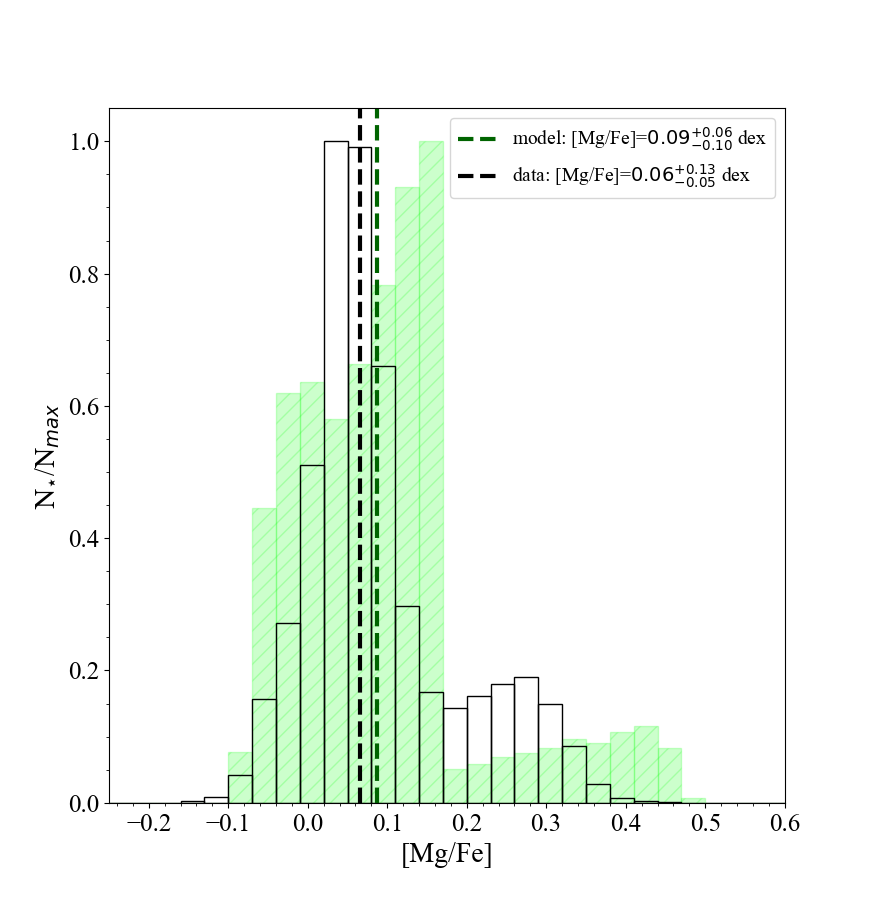}
\includegraphics[scale=0.24]{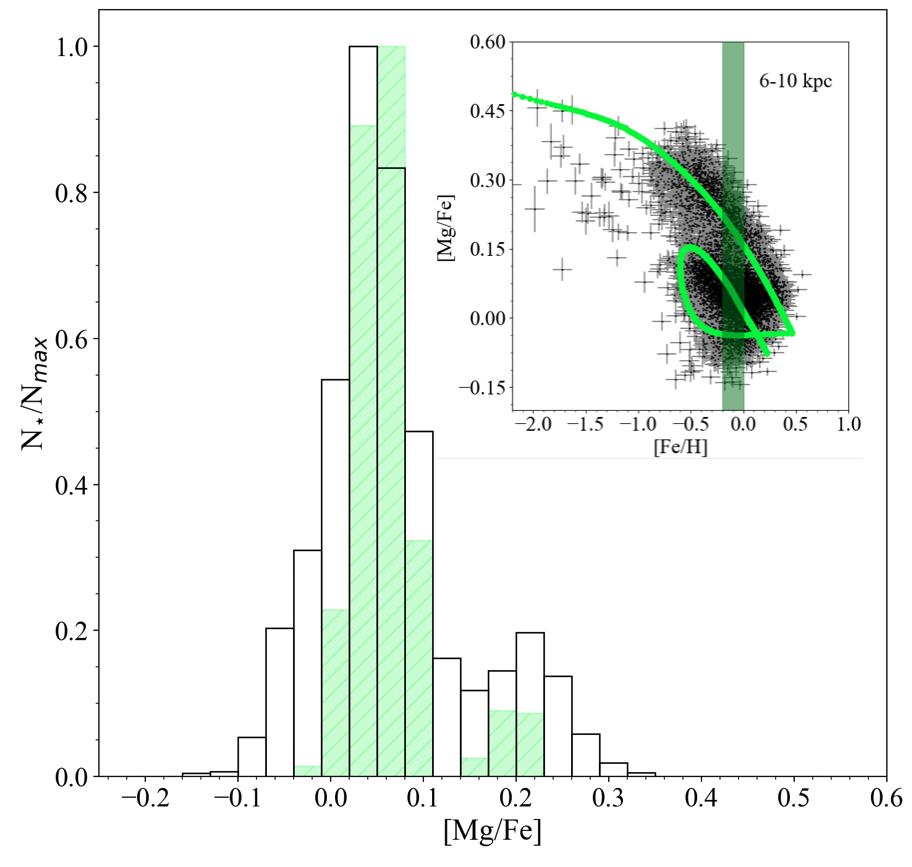}
\caption{{\it Left panel}: [Mg/Fe] distribution predicted by our best-fit model computed at 8 kpc (green histogram) compared with the APOGEE data in stars with Galactocentric distances between 6 and 10 kpc. Black and green vertical dashed lines indicate the median values of the data and model, respectively.  Distributions are normalised to the corresponding maximum number of stars, $N_{max}$.  {\it Right panel}:  Same as the left panel, but computed for a limited metallicity range, $-0.2 \, {\rm dex} \leq$ [Fe/H] $\leq 0  \, {\rm dex}$, as highlighted by the green shaded area in the inset.}
\label{MGFED_8}
\end{centering}
\end{figure*}

 In the left panel of Fig. \ref{SFR}  we show that the predicted present-day SFR   profile is in agreement with the observations by \citet{rana1991} and the  following analytical fit of SN remnants compilation by \citet{green2014}:
  
\begin{equation}
    {\rm SFR}(R)/{\rm SFR}_\odot= \bigg(\frac{R}{R_0}\bigg)^{b}\, e^{-c\big(\frac{R-R_0}{R_0}\big)},
\end{equation}
where $R_0=8$ kpc, $b=2$ and $c=5.1$ (see \citealt{palla2020} for more details).
 In the temporal evolution of the predicted SFR at different Galactocentric distances reported in the right panel of Fig. \ref{SFR},  the two  star formation phases (high-$\alpha$  and low-$\alpha$ stars)  and  the hiatus in between are evident. 
  The delay  $t_{max}$ between the two infall gas episodes is longer as we move towards the inner region. Such a  variation    is statistically significant (see Table \ref{tab_mcmc}),  and $t_{max}$ values span the range between 3.0 and 4.7 Gyr.  
Longer cooling timescales due to a more intense star formation activity and stronger  feedback are expected for the innermost Galactic regions at early times.
 Hence, we propose that  in the inner (outer) regions the  ISM gas  needed more (less) time to  cool down in order to  begin the SF activity associated with the low-$\alpha$ sequence, leading to larger (smaller) values for $t_{max}$.
This scenario is supported by several chemo-dynamical simulations in a cosmological framework where a hot gas phase is already in place at early times and the halo tends to inhibit gas filaments to penetrate into the central regions \citep{keres2005, dekel2006,brooks2009,fernandez2012, grand2018}.

 In Fig. \ref{SN}, we report  the time evolution of the Type Ia SN and Type II SN rates. The present-day Type II SN rate in the whole Galactic disc predicted by our model is 0.93 /[100 yr], a smaller value (but within 2$\sigma$ error) than the observations of \citet{li2010} which yield a value of 1.54 $\pm$0.32 /[100 yr]. 
The predicted present-day Type Ia SN rate in the whole Galactic disc
is 0.27 /[100 yr],  in good agreement with the value provided by \citet{cappellaro1997} of 0.30$\pm$0.20  /[100 yr]. In the lower panel of  Fig. \ref{SN} we also show the time evolution of the infall rate. We note that the present day value computed at 8 kpc is 1.01 M$\odot$ pc$^{-2}$ Gyr$^{-1}$,  consistent with the  range 0.3 -1.5 M$\odot$ pc$^{-2}$ Gyr$^{-1}$ suggested by \citet{matteucci2012}. 

In Fig. \ref{alpha_age} the time evolution of the metallicity [M/H] and the [$\alpha$/Fe] ratios computed at 4, 8, and 12 kpc are compared with the local   APOKASC sample by \citet{victor2018}. 
 Here, $\alpha$  is computed by means of  the sum of the abundances of Mg and Si.
The  metallicity [M/H] is computed, as in \citet{victor2018}, using the  following  expression introduced by \citet{salaris1993}:
\begin{equation} 
 \mbox{[M/H]}   = \mbox{[Fe/H]}   + \log \left( 0.638 \times
   10^{[\alpha/Fe]}+ 0.362 \right).
\label{MH}
\end{equation}
We combine the abundance ratios [Fe/H] and [$\alpha$/Fe] predicted by our model using this formulation to be consistent with the data. 
We notice that the best fit model at 8 kpc is very similar to the best model in   \citet{spitoni2019} constrained by APOKASC abundance ratios and asteroseismic ages.  \citet{spitoni2019} showed that the steep drop in [M/H] and bump in [$\alpha$/Fe] associated with the second accretion episode  (not obvious  in the observations),  are  hidden behind the observational uncertainties.

The highest metallicity  values are reached at any Galactic time by the innermost region.
Different slopes in the [M/H] and   [$\alpha$/Fe] ratios characterize the evolution of the low-$\alpha$ sequences at different Galactcocentric distances.  This is due to the interplay  of  different best-fit model  values for time-scale of accretion $\tau_2$,  SFEs, and gas infall enrichment  in diverse Galactic regions.

\begin{figure}
\begin{centering}
\includegraphics[scale=0.35]{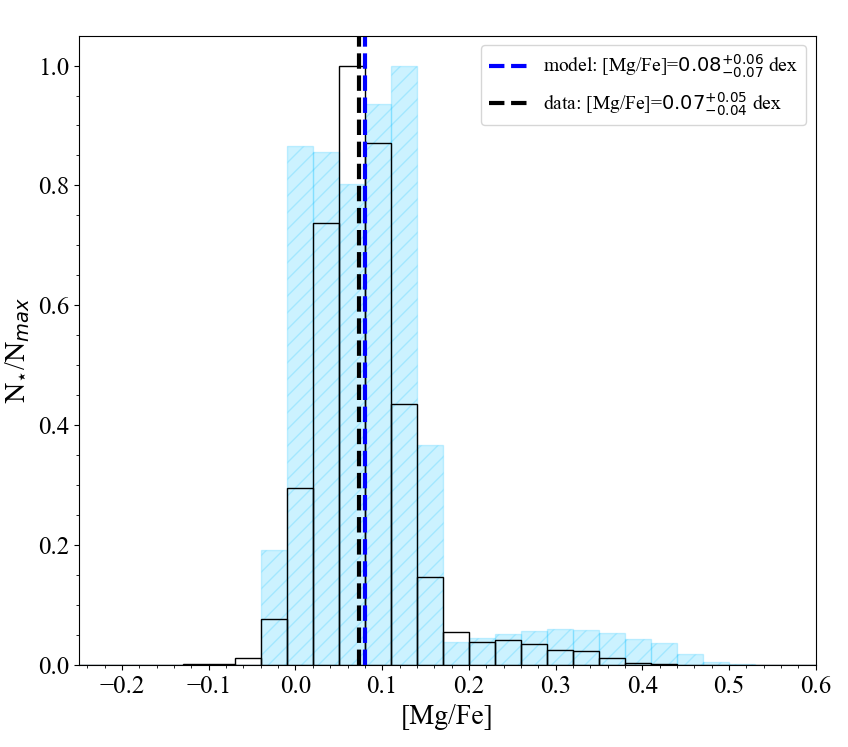}
\caption{[Mg/Fe] distribution predicted by our best-fit model computed at 12 kpc (blue histogram) compared with the APOGEE data in stars with Galactocentric distances between 10 and 14 kpc. Black and blue vertical lines indicate the median values of the data and model, respectively.}
\label{MGFED_12}
\end{centering}
\end{figure}

\begin{figure}
\begin{centering}
\includegraphics[scale=0.35]{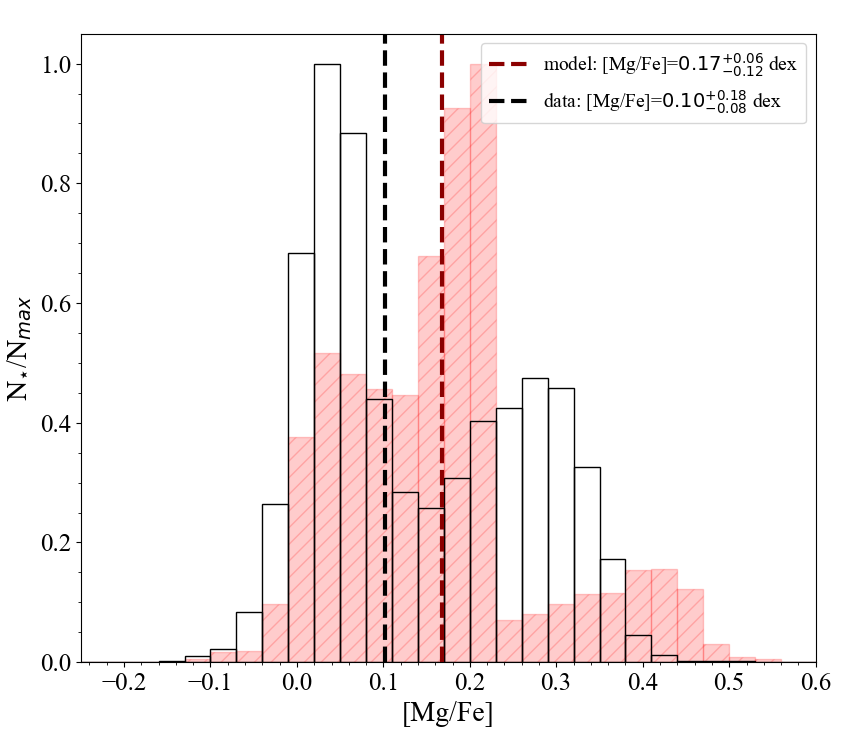}
\caption{[Mg/Fe] distribution predicted by our best-fit model computed at 4 kpc (red histogram) compared with the APOGEE data in a stellar sample with Galactocentric distances between 2 and 6 kpc. Black and red vertical lines indicate the median values of the data and model, respectively.}
\label{MGFED_4}
\end{centering}
\end{figure}

\begin{figure}
\begin{centering}
\includegraphics[scale=0.35]{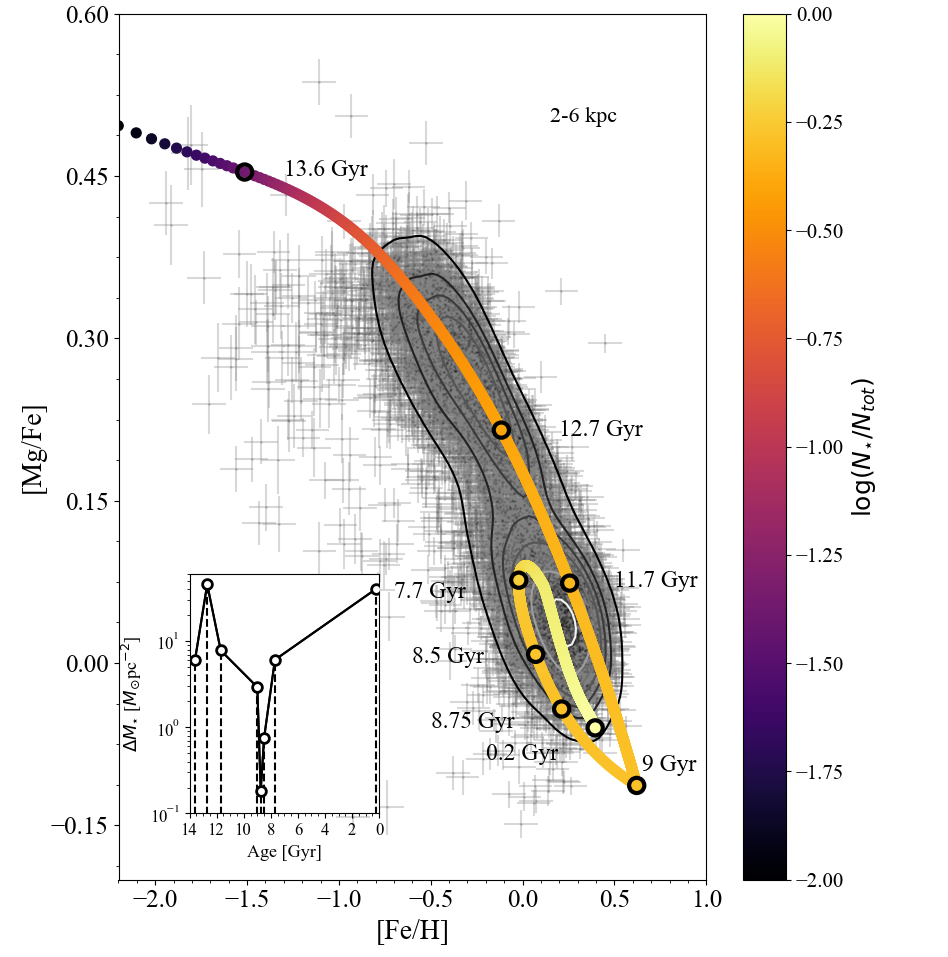}
\includegraphics[scale=0.33]{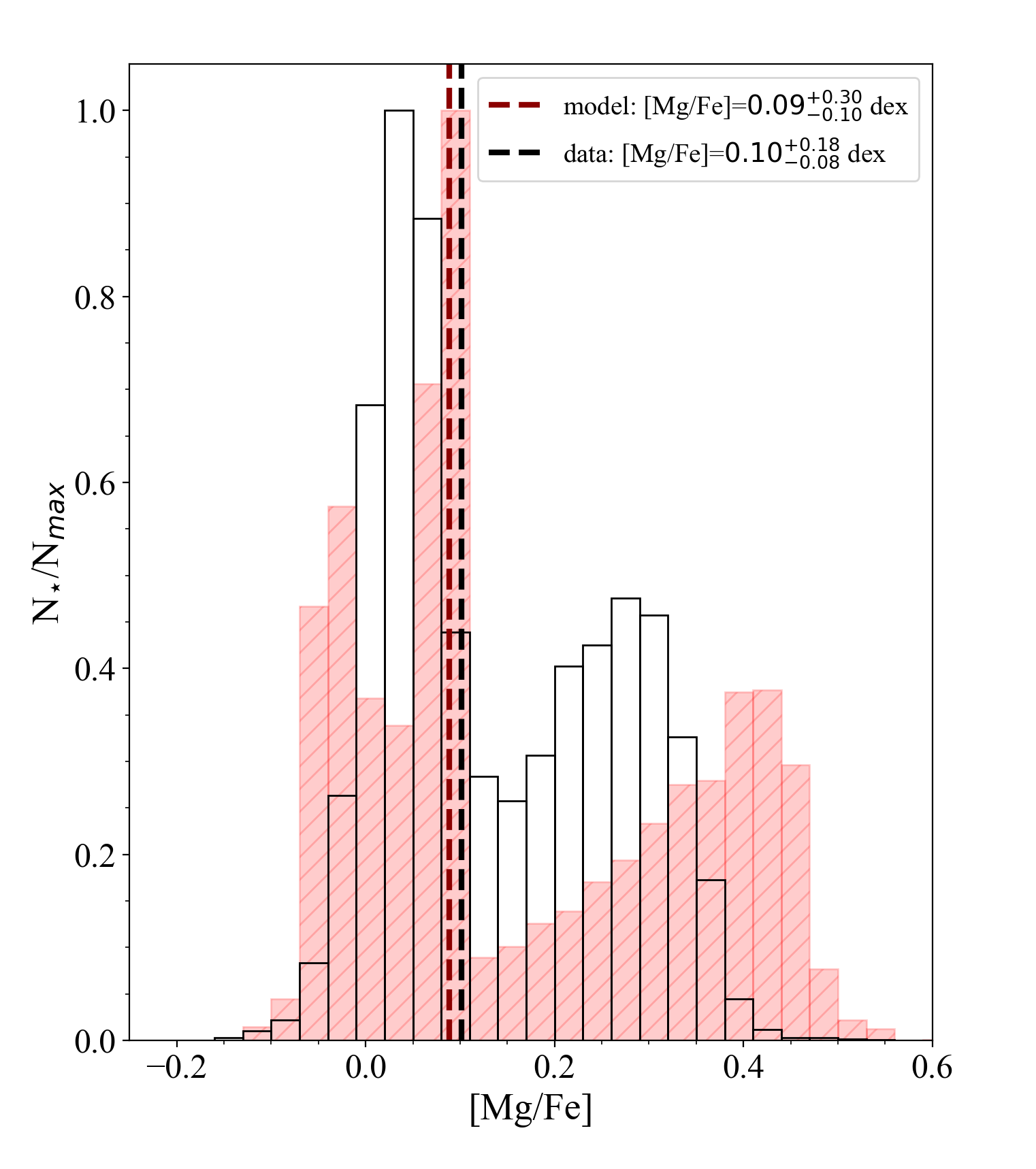}
\caption{APOGEE DR16 data in the region 2-6 kpc compared with model predictions  computed at 4 kpc with the same best-fit parameters as in the first column of Table \ref{tab_mcmc} but with the surface mass density ratio $\sigma_2/\sigma_1$ fixed at the value of 1.  {\it Upper panel}:
[Mg/Fe] vs. [Fe/H] abundance ratios. As in Fig. \ref{apogee_data}, the contour lines enclose fractions of 0.90, 0.75, 0.60, 0.45, 0.30, 0.20, 0.05 of the total number of observed stars.
The color coding represents the cumulative number of stars formed  during the Galactic evolution normalized to the total number $N_{tot}$.
%(the time-step is constant and fixed at the value of $\Delta \, t=4.6 \cdot 10^{-3}$ Gyr).
The open circles mark the model abundance ratios of stellar populations with different ages. 
In the inset we show the surface 
stellar mass density $\Delta M_{\star}$ formed in different age bins as a function of age, where the bin sizes are delimited by the vertical dashed lines and correspond to the same age values as indicated in the [Mg/Fe] vs [Fe/H] plot. 
 {\it Lower panel}: [Mg/Fe] distributions. Black and red vertical lines represent the median values of the data and model, respectively.}
\label{4s1}
\end{centering}
\end{figure}

\subsection{Metallicity  and [Mg/Fe] distribution functions}\label{sec_MDF}

In Fig. \ref{MDF}, it is shown that the predicted [Fe/H] distribution functions  at different Galactocentric distances are  generally in agreement with the data.

To highlight once more the low-$\alpha$ and high-$\alpha$ bimodality, in  \ref{MGFED_8}  we show 
 the [Mg/Fe] distributions, where we see that the APOGEE DR16 data
in the annular region centered at 8 kpc exhibit two neat peaks.  Although the best-fit model accounts for the observed  bimodality   and the  median distribution value   is consistent with the data, the  predicted peaks are shifted towards higher  [Mg/Fe] values. 
In the right panel of Fig.  \ref{MGFED_8},  we draw the same distribution but only for stars with
[Fe/H] in the range between -0.2 and 0 dex (see the highlighted region in the enclosed plot). In this case, the predictions are in better agreement with the data. 

The reason why the full data set seems in contrast with model is largely due to the large uncertainty in the assumed stellar nucleosynthesis yields of Mg from massive stars, which cause the model to have higher [Mg/Fe] ratios   ($\sim$0.45 dex) at low [Fe/H] than the observations from APOGEE.  Moreover, our assumption of a bottom-heavy IMF  \citep{scalo1986}  slows down the time evolution of the [Fe/H] abundances in the ISM, creating a bias towards higher [Mg/Fe] ratios after each infall event. 
  Additionally, since two-infall model is an approximate representation of the truth, significantly larger number of stars in the low-alpha sequence compared to the high-alpha can compromise the fit for the high-$\alpha$ sequence (as it gets less weight in the optimization process presented in Section \ref{fitting}).  
A better agreement is achieved for the external region centered at 12 kpc as shown in Fig. \ref{MGFED_12}, where the position of two peaks and the median value of the distribution are in good agreement with the data.

In Fig. \ref{MGFED_4} the observed and predicted [Mg/Fe] distribution  for the region centered at 4 kpc are compared.  Once again, we  see that the median value and the location of the two peaks are quite different from the data.
In order to understand this discrepancy,  we ran a new model with the same best-fit parameters as in Table \ref{tab_mcmc} for the 4 kpc case,  but with a smaller   surface mass density $\sigma_2/\sigma_1$ ratio. In   Fig. \ref{4s1}, we show the predicted  [Mg/Fe] vs. [Fe/H] relation  and the [Mg/Fe] distribution function   imposing   $\sigma_2/\sigma_1=1$. In  the upper panel of Fig. \ref{4s1}, we note that, as expected, the maximum [Mg/Fe] value reached  in the low-$\alpha$ sequence is smaller compared  to the one in Fig. \ref{mg4_SFE3} (with  $\sigma_2/\sigma_1=3.8$) because of the smaller mass associated  with the low-$\alpha$ phase as clearly shown in the inset plots of  Figs. \ref{mg4_SFE3} and \ref{4s1}.   In fact, the model with the lowest  $\sigma_2/\sigma_1$ ratio presents the highest increase of stellar mass $\Delta M_{\star}$ in the high-$\alpha$ phase.

In the lower panel of   Fig. \ref{4s1}, we can appreciate that the model reproduces  better the data than the results reported in Fig. \ref{MGFED_4} given also the intrinsic uncertainties due to the Mg stellar yields from massive stars and IMF. In fact,  the median values of the predicted [Mg/Fe] and observed data are pretty similar. In conclusion, a smaller $\sigma_2/\sigma_1$ has a double effect: (i) the peak of the [Mg/Fe] abundance associated with the low-$\alpha$ is shifted towards smaller values, and (ii) an increase of the number of the high-$\alpha$ stars (the second peak in the  distribution has a  higher number of stars  with larger [Mg/Fe] values than in Fig. \ref{MGFED_4}).

\subsection{Comparison with \citet{palla2020}} \label{palla}

Recently, \citet{palla2020} presented a revised Galactic chemical evolution model for  the  disc formation based on  the two-infall scenario.
In agreement with our findings, their best model suggests that  a variable SFE should be
acting together with the 'inside-out' mechanism for the low-$\alpha$ disc formation.
In addition, they claimed that radial gas inflows can help to create an abundance gradient  confirming the findings of   \citet{spitoni2015} and \citet{grisoni2018}.

In order to reproduce  the observed [Mg/Fe] vs. [Fe/H] of APOGEE \citep{hayden2015} at different Galactocentric distances, \citet{palla2020} proposed a delay  of $t_{max}$= 3.25 Gyr  between the two gas infall events, in agreement with the work presented here  and \citet{spitoni2019,spitoni2020}. 
In particular, they invoked an  enriched gas infall to properly reproduce the inner disc  [Mg/Fe] vs. [Fe/H] abundance ratio, in accordance  with our study (see Section \ref{sec_in}).

Differences in the chemical evolution tracks in the innermost disc region are primarily due   to  the different stellar nucleosynthesis prescriptions. Here, we adopt the yield collection proposed by \citet{francois2004}, whereas \citet{palla2020}  used those from \citet{romano2010}.
Moreover, our  results  are based on a  Bayesian analysis to fit the latest APOGEE DR16 data \citep{Ahumada2019}.
Finally,  \citet{palla2020} impose a different length-scale for the two disc components instead of a variable ratio between the low-$\alpha$ and high-$\alpha$ surface mass densities.
Notwithstanding all these differences, our model and the one of  \citet{palla2020}  share a similar  growth of the Galactic disc following the 'inside-out' scenario and    predict  pretty similar metallicity distribution functions.

\section{Conclusions}\label{conc}

We have presented a multi-zone chemical evolution model designed for the whole Galactic disc  constrained by  chemical abundance ratios of APOGEE DR16 \citep{Ahumada2019} data 
   using the   Bayesian analysis presented in \citet{spitoni2020}.     
In this study, we have considered four free parameters: accretion time-scales $\tau_1$ and $\tau_2$, delay $t_{max}$ and present-day surface mass density ratio $\sigma_2/\sigma_1$.

Our main conclusions can be summarized as follows.
\begin{enumerate}

\item  The Bayesian analysis based on the recent APOGEE DR16 data \citep{Ahumada2019}  suggests the presence of   a significant delay time  between the two gas infall episodes  for the thick-disc and thin-disc formation in  all analyzed Galactocentric regions. We find that the best values for the delay times are in the range between 3 and 4.7 Gyr, confirming the findings of \citet{spitoni2019,spitoni2020} for the solar neighborhood  based on the APOKASC data.

\item 
An inside-out formation of the thin-disc of our Galaxy naturally emerges from the best fit of our multi-zone chemical-evolution model to APOGEE-DR16 data:
 inner  Galactic regions are assembled on shorter time-scales  than  external Galactic zones. Moreover, our best-fit model predicts larger $\sigma_2/\sigma_1$ (ratio of low-$\alpha$ to high-$\alpha$ surface mass densities) values towards outer Galactic regions (see Fig. \ref{inside}), in agreement with the fact that as we move towards external regions, the APOGEE DR16 data sample presents less and less stars in the high-$\alpha$ phase compared to the    low-$\alpha$ sequence  \citep{queiroz2020}.

\item In  outer disc regions with Galactocentric distances   $R>6$ kpc, the chemical dilution   originating from a late gas accretion event with primordial chemical composition is the main driver of the [Mg/Fe] vs. [Fe/H] abundance pattern in the low-$\alpha$ phase,  extending the findings of the models presented by \citet{spitoni2019,spitoni2020} for  the solar neighborhood.

\item In the inner disc, for the  two-infall model to work, an enriched gas infall for the formation of low-$\alpha$ sequence stars  is required to  reproduce the observed data as suggested by \citet{palla2020}.  Different physical explanations could be invoked: the gas might be   enriched with metals from   outflows originated in massive galaxies \citep{VINTERGATANI2020,VINTERGATANII2020,VINTERGATANIII2020}, or it could be due  to  gas lost from the formation of the thick  disc,    which  then  gets mixed with a larger amount of infalling primordial gas \citep{gilmore1986}.

\item Our model   reproduces important observational constraints for the chemical evolution of the  whole disc reasonably well, such as  the present-day profiles of  the SFR, the stellar and gas surface densities.  Moreover, the predicted abundance gradient is in good agreement with the observations, thanks to the longer time-scales of accretion in the outer regions and to  a variable SFE for the low-$\alpha$ sequence. In the solar neighborhood, the model is able to reproduce the solar photospheric abundance values of \citet{asplund2005}.

\end{enumerate}

The above mentioned results  suggest that  the signatures of a delayed gas-rich merger giving rise to  a hiatus in the star formation history 
are impressed  in   the  [Mg/Fe] vs. [Fe/H] relation and determine  the distribution of the low-$\alpha$ stars in the abundance space at different  Galactocentric distances.

\section*{Acknowledgement}

Funding for the Stellar Astrophysics Centre is provided by The Danish National Research Foundation (Grant agreement no.: DNRF106).
E. Spitoni and V. Silva Aguirre acknowledge support from the Independent Research Fund Denmark (Research grant 7027-00096B). 
F. Vincenzo acknowledges the support of a Fellowship from the Center for Cosmology and AstroParticle Physics at The Ohio State University.
V. Grisoni acknowledges financial support at SISSA from the European Social Fund operational Programme 2014/2020 of the autonomous region Friuli Venezia Giulia. F. Calura  acknowledges support from grant PRIN MIUR
2017 - 20173ML3WW 001 and from the INAF main-stream
(1.05.01.86.31).

In this work, we have made use of SDSS-IV APOGEE-2 DR16 data. Funding for the Sloan Digital Sky Survey IV has been provided by the Alfred P. Sloan Foundation, the U.S. Department of Energy Office of Science, and the Participating Institutions. SDSS-IV acknowledges
support and resources from the Center for High-Performance Computing at
the University of Utah. The SDSS web site is  \href{www.sdss.org}{www.sdss.org}.
SDSS is managed by the Astrophysical Research Consortium for the Participating Institutions of the SDSS Collaboration which are listed at \href{https://www.sdss.org/collaboration/affiliations/}{www.sdss.org/collaboration/affiliations/}.

 This work has made use of data from the European Space Agency (ESA) mission
{\it Gaia} (\url{https://www.cosmos.esa.int/gaia}), processed by the {\it Gaia}
Data Processing and Analysis Consortium (DPAC,
\url{https://www.cosmos.esa.int/web/gaia/dpac/consortium}). Funding for the DPAC
has been provided by national institutions, in particular the institutions
participating in the {\it Gaia} Multilateral Agreement.
\bibliographystyle{aa} % style aa.bst
\bibliography{disk}

\begin{thebibliography}{116}
\expandafter\ifx\csname natexlab\endcsname\relax\def\natexlab#1{#1}\fi

\bibitem[{{Adibekyan} {et~al.}(2012){Adibekyan}, {Sousa}, {Santos}, {Delgado
  Mena}, {Gonz{\'a}lez Hern{\'a}ndez}, {Israelian}, {Mayor}, \&
  {Khachatryan}}]{adi2012}
{Adibekyan}, V.~Z., {Sousa}, S.~G., {Santos}, N.~C., {et~al.} 2012, \aap, 545,
  A32

\bibitem[{{Agertz} {et~al.}(2020){Agertz}, {Renaud}, {Feltzing}, {Read},
  {Ryde}, {Andersson}, {Rey}, {Bensby}, \& {Feuillet}}]{VINTERGATANI2020}
{Agertz}, O., {Renaud}, F., {Feltzing}, S., {et~al.} 2020, arXiv e-prints,
  arXiv:2006.06008

\bibitem[{{Ahumada} {et~al.}(2020){Ahumada}, {Allende Prieto}, {Almeida},
  {Anders}, {Anderson}, {Andrews}, {Anguiano}, {Arcodia}, {Armengaud},
  {Aubert}, {Avila}, {Avila-Reese}, {Badenes}, {Balland}, {Barger},
  {Barrera-Ballesteros}, {Basu}, {Bautista}, {Beaton}, {Beers}, {Benavides},
  {Bender}, {Bernardi}, {Bershady}, {Beutler}, {Bidin}, {Bird}, {Bizyaev},
  {Blanc}, {Blanton}, {Boquien}, {Borissova}, {Bovy}, {Brandt}, {Brinkmann},
  {Brownstein}, {Bundy}, {Bureau}, {Burgasser}, {Burtin}, {Cano-D{\'\i}az},
  {Capasso}, {Cappellari}, {Carrera}, {Chabanier}, {Chaplin}, {Chapman},
  {Cherinka}, {Chiappini}, {Doohyun Choi}, {Chojnowski}, {Chung}, {Clerc},
  {Coffey}, {Comerford}, {Comparat}, {da Costa}, {Cousinou}, {Covey}, {Crane},
  {Cunha}, {da Silva Ilha}, {Dai}, {Damsted}, {Darling}, {Davidson}, {Davies},
  {Dawson}, {De}, {de la Macorra}, {De Lee}, {de Andrade Queiroz}, {Deconto
  Machado}, {de la Torre}, {Dell'Agli}, {du Mas des Bourboux},
  {Diamond-Stanic}, {Dillon}, {Donor}, {Drory}, {Duckworth}, {Dwelly},
  {Ebelke}, {Eftekharzadeh}, {Eigenbrot}, {Elsworth}, {Eracleous},
  {Erfanianfar}, {Escoffier}, {Fan}, {Farr}, {Fern{\'a}ndez-Trincado},
  {Feuillet}, {Finoguenov}, {Fofie}, {Fraser-McKelvie}, {Frinchaboy},
  {Fromenteau}, {Fu}, {Galbany}, {Garcia}, {Garc{\'\i}a-Hern{\'a}ndez}, {Garma
  Oehmichen}, {Ge}, {Geimba Maia}, {Geisler}, {Gelfand}, {Goddy},
  {Gonzalez-Perez}, {Grabowski}, {Green}, {Grier}, {Guo}, {Guy}, {Harding},
  {Hasselquist}, {Hawken}, {Hayes}, {Hearty}, {Hekker}, {Hogg}, {Holtzman},
  {Horta}, {Hou}, {Hsieh}, {Huber}, {Hunt}, {Ider Chitham}, {Imig}, {Jaber},
  {Jimenez Angel}, {Johnson}, {Jones}, {J{\"o}nsson}, {Jullo}, {Kim},
  {Kinemuchi}, {Kirkpatrick}, {Kite}, {Klaene}, {Kneib}, {Kollmeier}, {Kong},
  {Kounkel}, {Krishnarao}, {Lacerna}, {Lan}, {Lane}, {Law}, {Le Goff}, {Leung},
  {Lewis}, {Li}, {Lian}, {Lin}, {Long}, {Longa-Pe{\~n}a}, {Lundgren}, {Lyke},
  {Ted Mackereth}, {MacLeod}, {Majewski}, {Manchado}, {Maraston}, {Martini},
  {Masseron}, {Masters}, {Mathur}, {McDermid}, {Merloni}, {Merrifield},
  {M{\'e}sz{\'a}ros}, {Miglio}, {Minniti}, {Minsley}, {Miyaji}, {Mohammad},
  {Mosser}, {Mueller}, {Muna}, {Mu{\~n}oz-Guti{\'e}rrez}, {Myers}, {Nadathur},
  {Nair}, {Nandra}, {do Nascimento}, {Nevin}, {Newman}, {Nidever}, {Nitschelm},
  {Noterdaeme}, {O'Connell}, {Olmstead}, {Oravetz}, {Oravetz}, {Osorio},
  {Pace}, {Padilla}, {Palanque-Delabrouille}, {Palicio}, {Pan}, {Pan},
  {Parker}, {Paviot}, {Peirani}, {Pe{\~n}a Ram{\'r}ez}, {Penny}, {Percival},
  {Perez-Fournon}, {P{\'e}rez-R{\`a}fols}, {Petitjean}, {Pieri},
  {Pinsonneault}, {Poovelil}, {Povick}, {Prakash}, {Price-Whelan}, {Raddick},
  {Raichoor}, {Ray}, {Rembold}, {Rezaie}, {Riffel}, {Riffel}, {Rix}, {Robin},
  {Roman-Lopes}, {Rom{\'a}n-Z{\'u}{\~n}iga}, {Rose}, {Ross}, {Rossi},
  {Rowlands}, {Rubin}, {Salvato}, {S{\'a}nchez}, {S{\'a}nchez-Menguiano},
  {S{\'a}nchez-Gallego}, {Sayres}, {Schaefer}, {Schiavon}, {Schimoia},
  {Schlafly}, {Schlegel}, {Schneider}, {Schultheis}, {Schwope}, {Seo},
  {Serenelli}, {Shafieloo}, {Shamsi}, {Shao}, {Shen}, {Shetrone}, {Shirley},
  {Silva Aguirre}, {Simon}, {Skrutskie}, {Slosar}, {Smethurst}, {Sobeck},
  {Sodi}, {Souto}, {Stark}, {Stassun}, {Steinmetz}, {Stello}, {Stermer},
  {Storchi-Bergmann}, {Streblyanska}, {Stringfellow}, {Stutz}, {Su{\'a}rez},
  {Sun}, {Taghizadeh-Popp}, {Talbot}, {Tayar}, {Thakar}, {Theriault}, {Thomas},
  {Thomas}, {Tinker}, {Tojeiro}, {Toledo}, {Tremonti}, {Troup}, {Tuttle},
  {Unda-Sanzana}, {Valentini}, {Vargas-Gonz{\'a}lez}, {Vargas-Maga{\~n}a},
  {V{\'a}zquez-Mata}, {Vivek}, {Wake}, {Wang}, {Weaver}, {Weijmans}, {Wild},
  {Wilson}, {Wilson}, {Wolthuis}, {Wood-Vasey}, {Yan}, {Yang}, {Y{\`e}che},
  {Zamora}, {Zarrouk}, {Zasowski}, {Zhang}, {Zhao}, {Zhao}, {Zheng}, {Zheng},
  {Zhu}, \& {Zou}}]{Ahumada2019}
{Ahumada}, R., {Allende Prieto}, C., {Almeida}, A., {et~al.} 2020, \apjs, 249,
  3

\bibitem[{{Asplund} {et~al.}(2005){Asplund}, {Grevesse}, \&
  {Sauval}}]{asplund2005}
{Asplund}, M., {Grevesse}, N., \& {Sauval}, A.~J. 2005, in Astronomical Society
  of the Pacific Conference Series, Vol. 336, Cosmic Abundances as Records of
  Stellar Evolution and Nucleosynthesis, ed. T.~G. {Barnes}, III \& F.~N.
  {Bash}, 25

\bibitem[{{Bailer-Jones} {et~al.}(2018){Bailer-Jones}, {Rybizki}, {Fouesneau},
  {Mantelet}, \& {Andrae}}]{BJ2018}
{Bailer-Jones}, C.~A.~L., {Rybizki}, J., {Fouesneau}, M., {Mantelet}, G., \&
  {Andrae}, R. 2018, \aj, 156, 58

\bibitem[{{Belfiore} {et~al.}(2019){Belfiore}, {Vincenzo}, {Maiolino}, \&
  {Matteucci}}]{belfiore2019}
{Belfiore}, F., {Vincenzo}, F., {Maiolino}, R., \& {Matteucci}, F. 2019,
  \mnras, 487, 456

\bibitem[{{Bennett} \& {Bovy}(2019)}]{bennett2019}
{Bennett}, M. \& {Bovy}, J. 2019, \mnras, 482, 1417

\bibitem[{{Bonaparte} {et~al.}(2013){Bonaparte}, {Matteucci}, {Recchi},
  {Spitoni}, {Pipino}, \& {Grieco}}]{bonaparte2013}
{Bonaparte}, I., {Matteucci}, F., {Recchi}, S., {et~al.} 2013, \mnras, 435,
  2460

\bibitem[{{Bond} {et~al.}(1991){Bond}, {Cole}, {Efstathiou}, \&
  {Kaiser}}]{bond1991}
{Bond}, J.~R., {Cole}, S., {Efstathiou}, G., \& {Kaiser}, N. 1991, \apj, 379,
  440

\bibitem[{{Brooks} {et~al.}(2009){Brooks}, {Governato}, {Quinn}, {Brook}, \&
  {Wadsley}}]{brooks2009}
{Brooks}, A.~M., {Governato}, F., {Quinn}, T., {Brook}, C.~B., \& {Wadsley}, J.
  2009, \apj, 694, 396

\bibitem[{{Buck}(2020)}]{buck2020}
{Buck}, T. 2020, \mnras, 491, 5435

\bibitem[{{Buder} {et~al.}(2019){Buder}, {Lind}, {Ness}, {Asplund}, {Duong},
  {Lin}, {Kos}, {Casagrande}, {Casey}, {Bland-Hawthorn}, {de Silva}, {D'Orazi},
  {Freeman}, {Martell}, {Schlesinger}, {Sharma}, {Simpson}, {Zucker},
  {Zwitter}, {{\v{C}}otar}, {Dotter}, {Hayden}, {Hyde}, {Kafle}, {Lewis},
  {Nataf}, {Nordlander}, {Reid}, {Rix}, {Sk{\'u}lad{\'o}ttir}, {Stello},
  {Ting}, {Traven}, {Wyse}, \& {Galah Collaboration}}]{buder2019}
{Buder}, S., {Lind}, K., {Ness}, M.~K., {et~al.} 2019, \aap, 624, A19

\bibitem[{{Calura} \& {Menci}(2009)}]{calura2009}
{Calura}, F. \& {Menci}, N. 2009, \mnras, 400, 1347

\bibitem[{{Cappellaro} \& {Turatto}(1997)}]{cappellaro1997}
{Cappellaro}, E. \& {Turatto}, M. 1997, in NATO Advanced Science Institutes
  (ASI) Series C, Vol. 486, NATO Advanced Science Institutes (ASI) Series C,
  ed. P.~{Ruiz-Lapuente}, R.~{Canal}, \& J.~{Isern}, 77

\bibitem[{{Cescutti} {et~al.}(2007){Cescutti}, {Matteucci}, {Fran{\c c}ois}, \&
  {Chiappini}}]{cescutti2007}
{Cescutti}, G., {Matteucci}, F., {Fran{\c c}ois}, P., \& {Chiappini}, C. 2007,
  \aap, 462, 943

\bibitem[{{Chaplin} {et~al.}(2020){Chaplin}, {Serenelli}, {Miglio}, {Morel},
  {Mackereth}, {Vincenzo}, {Kjeldsen}, {Basu}, {Ball}, {Stokholm}, {Verma},
  {Mosumgaard}, {Silva Aguirre}, {Mazumdar}, {Ranadive}, {Antia}, {Lebreton},
  {Ong}, {Appourchaux}, {Bedding}, {Christensen-Dalsgaard}, {Creevey},
  {Garc{\'\i}a}, {Handberg}, {Huber}, {Kawaler}, {Lund}, {Metcalfe}, {Stassun},
  {Bazot}, {Beck}, {Bell}, {Bergemann}, {Buzasi}, {Benomar}, {Bossini},
  {Bugnet}, {Campante}, {Orhan}, {Corsaro}, {Gonz{\'a}lez-Cuesta}, {Davies},
  {Di Mauro}, {Egeland}, {Elsworth}, {Gaulme}, {Ghasemi}, {Guo}, {Hall},
  {Hasanzadeh}, {Hekker}, {Howe}, {Jenkins}, {Jim{\'e}nez}, {Kiefer},
  {Kuszlewicz}, {Kallinger}, {Latham}, {Lundkvist}, {Mathur}, {Montalb{\'a}n},
  {Mosser}, {Bed{\'o}n}, {Nielsen}, {{\"O}rtel}, {Rendle}, {Ricker},
  {Rodrigues}, {Roxburgh}, {Safari}, {Schofield}, {Seager}, {Smalley},
  {Stello}, {Szab{\'o}}, {Tayar}, {Theme{\ss}l}, {Thomas}, {Vanderspek}, {van
  Rossem}, {Vrard}, {Weiss}, {White}, {Winn}, \& {Y{\i}ld{\i}z}}]{chaplin2020}
{Chaplin}, W.~J., {Serenelli}, A.~M., {Miglio}, A., {et~al.} 2020, Nature
  Astronomy, 4, 382

\bibitem[{{Chiappini} {et~al.}(1997){Chiappini}, {Matteucci}, \&
  {Gratton}}]{chiappini1997}
{Chiappini}, C., {Matteucci}, F., \& {Gratton}, R. 1997, \apj, 477, 765

\bibitem[{{Chiappini} {et~al.}(2001){Chiappini}, {Matteucci}, \&
  {Romano}}]{chiappini2001}
{Chiappini}, C., {Matteucci}, F., \& {Romano}, D. 2001, \apj, 554, 1044

\bibitem[{{Colavitti} {et~al.}(2008){Colavitti}, {Matteucci}, \&
  {Murante}}]{colavitti2008}
{Colavitti}, E., {Matteucci}, F., \& {Murante}, G. 2008, \aap, 483, 401

\bibitem[{{Cole} {et~al.}(2000){Cole}, {Lacey}, {Baugh}, \& {Frenk}}]{cole2000}
{Cole}, S., {Lacey}, C.~G., {Baugh}, C.~M., \& {Frenk}, C.~S. 2000, \mnras,
  319, 168

\bibitem[{{C{\^o}t{\'e}} {et~al.}(2017){C{\^o}t{\'e}}, {O'Shea}, {Ritter},
  {Herwig}, \& {Venn}}]{cote2017}
{C{\^o}t{\'e}}, B., {O'Shea}, B.~W., {Ritter}, C., {Herwig}, F., \& {Venn},
  K.~A. 2017, \apj, 835, 128

\bibitem[{{Dame}(1993)}]{dame1993}
{Dame}, T.~M. 1993, in American Institute of Physics Conference Series, Vol.
  278, Back to the Galaxy, ed. S.~S. {Holt} \& F.~{Verter}, 267--278

\bibitem[{{Dekel} \& {Birnboim}(2006)}]{dekel2006}
{Dekel}, A. \& {Birnboim}, Y. 2006, \mnras, 368, 2

\bibitem[{{Fern{\'a}ndez} {et~al.}(2012){Fern{\'a}ndez}, {Joung}, \&
  {Putman}}]{fernandez2012}
{Fern{\'a}ndez}, X., {Joung}, M.~R., \& {Putman}, M.~E. 2012, \apj, 749, 181

\bibitem[{{Foreman-Mackey} {et~al.}(2013){Foreman-Mackey}, {Hogg}, {Lang}, \&
  {Goodman}}]{foreman}
{Foreman-Mackey}, D., {Hogg}, D.~W., {Lang}, D., \& {Goodman}, J. 2013, \pasp,
  125, 306

\bibitem[{{Fran{\c c}ois} {et~al.}(2004){Fran{\c c}ois}, {Matteucci}, {Cayrel},
  {Spite}, {Spite}, \& {Chiappini}}]{francois2004}
{Fran{\c c}ois}, P., {Matteucci}, F., {Cayrel}, R., {et~al.} 2004, \aap, 421,
  613

\bibitem[{{Frankel} {et~al.}(2018){Frankel}, {Rix}, {Ting}, {Ness}, \&
  {Hogg}}]{frankel2018}
{Frankel}, N., {Rix}, H.-W., {Ting}, Y.-S., {Ness}, M., \& {Hogg}, D.~W. 2018,
  \apj, 865, 96

\bibitem[{{Frankel} {et~al.}(2019){Frankel}, {Sanders}, {Rix}, {Ting}, \&
  {Ness}}]{frankel2019}
{Frankel}, N., {Sanders}, J., {Rix}, H.-W., {Ting}, Y.-S., \& {Ness}, M. 2019,
  \apj, 884, 99

\bibitem[{{Fuhrmann} {et~al.}(2017){Fuhrmann}, {Chini}, {Kaderhandt}, \&
  {Chen}}]{fuhr2017}
{Fuhrmann}, K., {Chini}, R., {Kaderhandt}, L., \& {Chen}, Z. 2017, \mnras, 464,
  2610

\bibitem[{{Gaia Collaboration} {et~al.}(2018){Gaia Collaboration}, {Katz},
  {Antoja}, {Romero-G{\'o}mez}, {Drimmel}, {Reyl{\'e}}, {Seabroke}, {Soubiran},
  {Babusiaux}, {Di Matteo}, {Figueras}, {Poggio}, {Robin}, {Evans}, {Brown},
  {Vallenari}, {Prusti}, {de Bruijne}, {Bailer-Jones}, {Biermann}, {Eyer},
  {Jansen}, {Jordi}, {Klioner}, {Lammers}, {Lindegren}, {Luri}, {Mignard},
  {Panem}, {Pourbaix}, {Randich}, {Sartoretti}, {Siddiqui}, {van Leeuwen},
  {Walton}, {Arenou}, {Bastian}, {Cropper}, {Lattanzi}, {Bakker}, {Cacciari},
  {Casta n}, {Chaoul}, {Cheek}, {De Angeli}, {Fabricius}, {Guerra}, {Holl},
  {Masana}, {Messineo}, {Mowlavi}, {Nienartowicz}, {Panuzzo}, {Portell},
  {Riello}, {Tanga}, {Th{\'e}venin}, {Gracia-Abril}, {Comoretto},
  {Garcia-Reinaldos}, {Teyssier}, {Altmann}, {Andrae}, {Audard},
  {Bellas-Velidis}, {Benson}, {Berthier}, {Blomme}, {Burgess}, {Busso},
  {Carry}, {Cellino}, {Clementini}, {Clotet}, {Creevey}, {Davidson}, {De
  Ridder}, {Delchambre}, {Dell'Oro}, {Ducourant},
  {Fern{\'a}ndez-Hern{\'a}ndez}, {Fouesneau}, {Fr{\'e}mat}, {Galluccio},
  {Garc{\'\i}a-Torres}, {Gonz{\'a}lez-N{\'u}{\~n}ez}, {Gonz{\'a}lez-Vidal},
  {Gosset}, {Guy}, {Halbwachs}, {Hambly}, {Harrison}, {Hern{\'a}ndez},
  {Hestroffer}, {Hodgkin}, {Hutton}, {Jasniewicz}, {Jean-Antoine-Piccolo},
  {Jordan}, {Korn}, {Krone-Martins}, {Lanzafame}, {Lebzelter}, {L{\"o}ffler},
  {Manteiga}, {Marrese}, {Mart{\'\i}n-Fleitas}, {Moitinho}, {Mora}, {Muinonen},
  {Osinde}, {Pancino}, {Pauwels}, {Petit}, {Recio-Blanco}, {Richards},
  {Rimoldini}, {Sarro}, {Siopis}, {Smith}, {Sozzetti}, {S{\"u}veges}, {Torra},
  {van Reeven}, {Abbas}, {Abreu Aramburu}, {Accart}, {Aerts}, {Altavilla},
  {{\'A}lvarez}, {Alvarez}, {Alves}, {Anderson}, {Andrei}, {Anglada Varela},
  {Antiche}, {Arcay}, {Astraatmadja}, {Bach}, {Baker},
  {Balaguer-N{\'u}{\~n}ez}, {Balm}, {Barache}, {Barata}, {Barbato}, {Barblan},
  {Barklem}, {Barrado}, {Barros}, {Barstow}, {Bartholom{\'e} Mu{\~n}oz},
  {Bassilana}, {Becciani}, {Bellazzini}, {Berihuete}, {Bertone}, {Bianchi},
  {Bienaym{\'e}}, {Blanco-Cuaresma}, {Boch}, {Boeche}, {Bombrun}, {Borrachero},
  {Bossini}, {Bouquillon}, {Bourda}, {Bragaglia}, {Bramante}, {Breddels},
  {Bressan}, {Brouillet}, {Br{\"u}semeister}, {Brugaletta}, {Bucciarelli},
  {Burlacu}, {Busonero}, {Butkevich}, {Buzzi}, {Caffau}, {Cancelliere},
  {Cannizzaro}, {Cantat-Gaudin}, {Carballo}, {Carlucci}, {Carrasco},
  {Casamiquela}, {Castellani}, {Castro-Ginard}, {Charlot}, {Chemin},
  {Chiavassa}, {Cocozza}, {Costigan}, {Cowell}, {Crifo}, {Crosta}, {Crowley},
  {Cuypers}, {Dafonte}, {Damerdji}, {Dapergolas}, {David}, {David}, {de
  Laverny}, {De Luise}, {De March}, {de Souza}, {de Torres}, {Debosscher}, {del
  Pozo}, {Delbo}, {Delgado}, {Delgado}, {Diakite}, {Diener}, {Distefano},
  {Dolding}, {Drazinos}, {Dur{\'a}n}, {Edvardsson}, {Enke}, {Eriksson},
  {Esquej}, {Eynard Bontemps}, {Fabre}, {Fabrizio}, {Faigler}, {Falc a},
  {Farr{\`a}s Casas}, {Federici}, {Fedorets}, {Fernique}, {Filippi},
  {Findeisen}, {Fonti}, {Fraile}, {Fraser}, {Fr{\'e}zouls}, {Gai}, {Galleti},
  {Garabato}, {Garc{\'\i}a-Sedano}, {Garofalo}, {Garralda}, {Gavel}, {Gavras},
  {Gerssen}, {Geyer}, {Giacobbe}, {Gilmore}, {Girona}, {Giuffrida}, {Glass},
  {Gomes}, {Granvik}, {Gueguen}, {Guerrier}, {Guiraud}, {Guti{\'e}}, {Haigron},
  {Hatzidimitriou}, {Hauser}, {Haywood}, {Heiter}, {Helmi}, {Heu}, {Hilger},
  {Hobbs}, {Hofmann}, {Holland }, {Huckle}, {Hypki}, {Icardi}, {Jan{\ss}en},
  {Jevardat de Fombelle}, {Jonker}, {Juh{\'a}sz}, {Julbe}, {Karampelas},
  {Kewley}, {Klar}, {Kochoska}, {Kohley}, {Kolenberg}, {Kontizas}, {Kontizas},
  {Koposov}, {Kordopatis}, {Kostrzewa-Rutkowska}, {Koubsky}, {Lambert},
  {Lanza}, {Lasne}, {Lavigne}, {Le Fustec}, {Le Poncin-Lafitte}, {Lebreton},
  {Leccia}, {Leclerc}, {Lecoeur-Taibi}, {Lenhardt}, {Leroux}, {Liao}, {Licata},
  {Lindstr{\o}m}, {Lister}, {Livanou}, {Lobel}, {L{\'o}pez}, {Managau}, {Mann},
  {Mantelet}, {Marchal}, {Marchant}, {Marconi}, {Marinoni}, {Marschalk{\'o}},
  {Marshall}, {Martino}, {Marton}, {Mary}, {Massari}, {Matijevi{\v{c}}},
  {Mazeh}, {McMillan}, {Messina}, {Michalik}, {Millar}, {Molina}, {Molinaro},
  {Moln{\'a}r}, {Montegriffo}, {Mor}, {Morbidelli}, {Morel}, {Morris},
  {Mulone}, {Muraveva}, {Musella}, {Nelemans}, {Nicastro}, {Noval},
  {O'Mullane}, {Ord{\'e}novic}, {Ord{\'o}{\~n}ez-Blanco}, {Osborne}, {Pagani},
  {Pagano}, {Pailler}, {Palacin}, {Palaversa}, {Panahi}, {Pawlak},
  {Piersimoni}, {Pineau}, {Plachy}, {Plum}, {Poujoulet}, {Pr{\v{s}}a},
  {Pulone}, {Racero}, {Ragaini}, {Rambaux}, {Ramos-Lerate}, {Regibo}, {Riclet},
  {Ripepi}, {Riva}, {Rivard}, {Rixon}, {Roegiers}, {Roelens}, {Rowell},
  {Royer}, {Ruiz-Dern}, {Sadowski}, {Sagrist{\`a} Sell{\'e}s}, {Sahlmann},
  {Salgado}, {Salguero}, {Sanna}, {Santana-Ros}, {Sarasso}, {Savietto},
  {Schultheis}, {Sciacca}, {Segol}, {Segovia}, {S{\'e}gransan}, {Shih},
  {Siltala}, {Silva}, {Smart}, {Smith}, {Solano}, {Solitro}, {Sordo}, {Soria
  Nieto}, {Souchay}, {Spagna}, {Spoto}, {Stampa}, {Steele},
  {Steidelm{\"u}ller}, {Stephenson}, {Stoev}, {Suess}, {Surdej}, {Szabados},
  {Szegedi-Elek}, {Tapiador}, {Taris}, {Tauran}, {Taylor}, {Teixeira},
  {Terrett}, {Teyssand ier}, {Thuillot}, {Titarenko}, {Torra Clotet}, {Turon},
  {Ulla}, {Utrilla}, {Uzzi}, {Vaillant}, {Valentini}, {Valette}, {van Elteren},
  {Van Hemelryck}, {van Leeuwen}, {Vaschetto}, {Vecchiato}, {Veljanoski},
  {Viala}, {Vicente}, {Vogt}, {von Essen}, {Voss}, {Votruba}, {Voutsinas},
  {Walmsley}, {Weiler}, {Wertz}, {Wevers}, {Wyrzykowski}, {Yoldas},
  {{\v{Z}}erjal}, {Ziaeepour}, {Zorec}, {Zschocke}, {Zucker}, {Zurbach}, \&
  {Zwitter}}]{gaia2_2018}
{Gaia Collaboration}, {Katz}, D., {Antoja}, T., {et~al.} 2018, \aap, 616, A11

\bibitem[{{Genovali} {et~al.}(2015){Genovali}, {Lemasle}, {da Silva}, {Bono},
  {Fabrizio}, {Bergemann}, {Buonanno}, {Ferraro}, {Fran{\c{c}}ois},
  {Iannicola}, {Inno}, {Laney}, {Kudritzki}, {Matsunaga}, {Nonino}, {Primas},
  {Romaniello}, {Urbaneja}, \& {Th{\'e}venin}}]{genovali2015}
{Genovali}, K., {Lemasle}, B., {da Silva}, R., {et~al.} 2015, \aap, 580, A17

\bibitem[{{Gilmore} \& {Wyse}(1986)}]{gilmore1986}
{Gilmore}, G. \& {Wyse}, R.~F.~G. 1986, \nat, 322, 806

\bibitem[{{Goodman} \& {Weare}(2010)}]{goodman2010}
{Goodman}, J. \& {Weare}, J. 2010, Communications in Applied Mathematics and
  Computational Science, 5, 65

\bibitem[{{Grand} {et~al.}(2018){Grand}, {Bustamante}, {G{\'o}mez}, {Kawata},
  {Marinacci}, {Pakmor}, {Rix}, {Simpson}, {Sparre}, \& {Springel}}]{grand2018}
{Grand}, R.~J.~J., {Bustamante}, S., {G{\'o}mez}, F.~A., {et~al.} 2018, \mnras,
  474, 3629

\bibitem[{{Gravity Collaboration} {et~al.}(2018){Gravity Collaboration},
  {Abuter}, {Amorim}, {Anugu}, {Baub{\"o}ck}, {Benisty}, {Berger}, {Blind},
  {Bonnet}, {Brandner}, {Buron}, {Collin}, {Chapron}, {Cl{\'e}net}, {Coud{\'e}
  Du Foresto}, {de Zeeuw}, {Deen}, {Delplancke-Str{\"o}bele}, {Dembet},
  {Dexter}, {Duvert}, {Eckart}, {Eisenhauer}, {Finger}, {F{\"o}rster
  Schreiber}, {F{\'e}dou}, {Garcia}, {Garcia Lopez}, {Gao}, {Gendron},
  {Genzel}, {Gillessen}, {Gordo}, {Habibi}, {Haubois}, {Haug}, {Hau{\ss}mann},
  {Henning}, {Hippler}, {Horrobin}, {Hubert}, {Hubin}, {Jimenez Rosales},
  {Jochum}, {Jocou}, {Kaufer}, {Kellner}, {Kendrew}, {Kervella}, {Kok},
  {Kulas}, {Lacour}, {Lapeyr{\`e}re}, {Lazareff}, {Le Bouquin}, {L{\'e}na},
  {Lippa}, {Lenzen}, {M{\'e}rand}, {M{\"u}ler}, {Neumann}, {Ott}, {Palanca},
  {Paumard}, {Pasquini}, {Perraut}, {Perrin}, {Pfuhl}, {Plewa}, {Rabien},
  {Ram{\'\i}rez}, {Ramos}, {Rau}, {Rodr{\'\i}guez-Coira}, {Rohloff}, {Rousset},
  {Sanchez-Bermudez}, {Scheithauer}, {Sch{\"o}ller}, {Schuler}, {Spyromilio},
  {Straub}, {Straubmeier}, {Sturm}, {Tacconi}, {Tristram}, {Vincent}, {von
  Fellenberg}, {Wank}, {Waisberg}, {Widmann}, {Wieprecht}, {Wiest},
  {Wiezorrek}, {Woillez}, {Yazici}, {Ziegler}, \& {Zins}}]{abuter2018}
{Gravity Collaboration}, {Abuter}, R., {Amorim}, A., {et~al.} 2018, \aap, 615,
  L15

\bibitem[{{Green}(2014)}]{green2014}
{Green}, D.~A. 2014, in IAU Symposium, Vol. 296, Supernova Environmental
  Impacts, ed. A.~{Ray} \& R.~A. {McCray}, 188--196

\bibitem[{{Grevesse} {et~al.}(1996){Grevesse}, {Noels}, \&
  {Sauval}}]{grevesse1996}
{Grevesse}, N., {Noels}, A., \& {Sauval}, A.~J. 1996, in Astronomical Society
  of the Pacific Conference Series, Vol.~99, Cosmic Abundances, ed. S.~S.
  {Holt} \& G.~{Sonneborn}, 117

\bibitem[{{Griffith} {et~al.}(2020){Griffith}, {Weinberg}, {Johnson}, {Beaton},
  {Garc{\'\i}a-Hern{\'a}ndez}, {Hasselquist}, {Holtzman}, {Johnson},
  {J{\"o}nsson}, {Lane}, {Nataf}, \& {Roman-Lopes}}]{griffith2020}
{Griffith}, E., {Weinberg}, D.~H., {Johnson}, J.~A., {et~al.} 2020, arXiv
  e-prints, arXiv:2009.05063

\bibitem[{{Grisoni} {et~al.}(2019){Grisoni}, {Matteucci}, {Romano}, \&
  {Fu}}]{grisoni2019}
{Grisoni}, V., {Matteucci}, F., {Romano}, D., \& {Fu}, X. 2019, \mnras, 489,
  3539

\bibitem[{{Grisoni} {et~al.}(2020){Grisoni}, {Romano}, {Spitoni}, {Matteucci},
  {Ryde}, \& {J{\"o}nsson}}]{grisoni2020}
{Grisoni}, V., {Romano}, D., {Spitoni}, E., {et~al.} 2020, \mnras, 498, 1252

\bibitem[{{Grisoni} {et~al.}(2018){Grisoni}, {Spitoni}, \&
  {Matteucci}}]{grisoni2018}
{Grisoni}, V., {Spitoni}, E., \& {Matteucci}, F. 2018, \mnras, 481, 2570

\bibitem[{{Grisoni} {et~al.}(2017){Grisoni}, {Spitoni}, {Matteucci},
  {Recio-Blanco}, {de Laverny}, {Hayden}, {Mikolaitis}, \&
  {Worley}}]{grisoni2017}
{Grisoni}, V., {Spitoni}, E., {Matteucci}, F., {et~al.} 2017, \mnras, 472, 3637

\bibitem[{{Hayden} {et~al.}(2015){Hayden}, {Bovy}, {Holtzman}, {Nidever},
  {Bird}, {Weinberg}, {Andrews}, {Majewski}, {Allende Prieto}, {Anders},
  {Beers}, {Bizyaev}, {Chiappini}, {Cunha}, {Frinchaboy},
  {Garc{\'{\i}}a-Her{\'n}andez}, {Garc{\'{\i}}a P{\'e}rez}, {Girardi},
  {Harding}, {Hearty}, {Johnson}, {M{\'e}sz{\'a}ros}, {Minchev}, {O'Connell},
  {Pan}, {Robin}, {Schiavon}, {Schneider}, {Schultheis}, {Shetrone},
  {Skrutskie}, {Steinmetz}, {Smith}, {Wilson}, {Zamora}, \&
  {Zasowski}}]{hayden2015}
{Hayden}, M.~R., {Bovy}, J., {Holtzman}, J.~A., {et~al.} 2015, \apj, 808, 132

\bibitem[{{Haywood} {et~al.}(2013){Haywood}, {Di Matteo}, {Lehnert}, {Katz}, \&
  {G{\'o}mez}}]{haywood2013}
{Haywood}, M., {Di Matteo}, P., {Lehnert}, M.~D., {Katz}, D., \& {G{\'o}mez},
  A. 2013, \aap, 560, A109

\bibitem[{{Helmi} {et~al.}(2018){Helmi}, {Babusiaux}, {Koppelman}, {Massari},
  {Veljanoski}, \& {Brown}}]{helmi2018}
{Helmi}, A., {Babusiaux}, C., {Koppelman}, H.~H., {et~al.} 2018, \nat, 563, 85

\bibitem[{{Iwamoto} {et~al.}(1999){Iwamoto}, {Brachwitz}, {Nomoto},
  {Kishimoto}, {Umeda}, {Hix}, \& {Thielemann}}]{iwamoto1999}
{Iwamoto}, K., {Brachwitz}, F., {Nomoto}, K., {et~al.} 1999, \apjs, 125, 439

\bibitem[{{Kennicutt}(1998)}]{kenni1998}
{Kennicutt}, Jr., R.~C. 1998, \apj, 498, 541

\bibitem[{{Kere{\v{s}}} {et~al.}(2005){Kere{\v{s}}}, {Katz}, {Weinberg}, \&
  {Dav{\'e}}}]{keres2005}
{Kere{\v{s}}}, D., {Katz}, N., {Weinberg}, D.~H., \& {Dav{\'e}}, R. 2005,
  \mnras, 363, 2

\bibitem[{{Khoperskov} {et~al.}(2021){Khoperskov}, {Haywood}, {Snaith}, {Di
  Matteo}, {Lehnert}, {Vasiliev}, {Naroenkov}, \& {Berczik}}]{Khoperskov2020}
{Khoperskov}, S., {Haywood}, M., {Snaith}, O., {et~al.} 2021, \mnras
  [\eprint[arXiv]{2006.10195}]

\bibitem[{{Koppelman} {et~al.}(2019){Koppelman}, {Helmi}, {Massari},
  {Price-Whelan}, \& {Starkenburg}}]{koppelman2019}
{Koppelman}, H.~H., {Helmi}, A., {Massari}, D., {Price-Whelan}, A.~M., \&
  {Starkenburg}, T.~K. 2019, \aap, 631, L9

\bibitem[{{Kubryk} {et~al.}(2013){Kubryk}, {Prantzos}, \&
  {Athanassoula}}]{kubryk2013}
{Kubryk}, M., {Prantzos}, N., \& {Athanassoula}, E. 2013, \mnras, 436, 1479

\bibitem[{{Kubryk} {et~al.}(2015){Kubryk}, {Prantzos}, \&
  {Athanassoula}}]{kubryk2015}
{Kubryk}, M., {Prantzos}, N., \& {Athanassoula}, E. 2015, \aap, 580, A126

\bibitem[{{Larson}(1976)}]{larson1976}
{Larson}, R.~B. 1976, \mnras, 176, 31

\bibitem[{{Leung} \& {Bovy}(2019)}]{leung2019}
{Leung}, H.~W. \& {Bovy}, J. 2019, \mnras, 489, 2079

\bibitem[{{Li} {et~al.}(2011){Li}, {Chornock}, {Leaman}, {Filippenko},
  {Poznanski}, {Wang}, {Ganeshalingam}, \& {Mannucci}}]{li2010}
{Li}, W., {Chornock}, R., {Leaman}, J., {et~al.} 2011, \mnras, 412, 1473

\bibitem[{{Lian} {et~al.}(2020){Lian}, {Thomas}, {Maraston}, {Beers}, {Moni
  Bidin}, {Fern{\'a}ndez-Trincado}, {Garc{\'\i}a-Hern{\'a}ndez}, {Lane},
  {Munoz}, {Nitschelm}, {Roman-Lopes}, \& {Zamora}}]{lian2020b}
{Lian}, J., {Thomas}, D., {Maraston}, C., {et~al.} 2020, \mnras, 497, 2371

\bibitem[{{Loebman} {et~al.}(2011){Loebman}, {Ro{\v{s}}kar}, {Debattista},
  {Ivezi{\'c}}, {Quinn}, \& {Wadsley}}]{loebmn2011}
{Loebman}, S.~R., {Ro{\v{s}}kar}, R., {Debattista}, V.~P., {et~al.} 2011, \apj,
  737, 8

\bibitem[{{Luck} \& {Lambert}(2011)}]{luck2011}
{Luck}, R.~E. \& {Lambert}, D.~L. 2011, \aj, 142, 136

\bibitem[{{Luri} {et~al.}(2018){Luri}, {Brown}, {Sarro}, {Arenou},
  {Bailer-Jones}, {Castro-Ginard}, {de Bruijne}, {Prusti}, {Babusiaux}, \&
  {Delgado}}]{luri2018}
{Luri}, X., {Brown}, A.~G.~A., {Sarro}, L.~M., {et~al.} 2018, \aap, 616, A9

\bibitem[{{Mackereth} {et~al.}(2017){Mackereth}, {Bovy}, {Schiavon},
  {Zasowski}, {Cunha}, {Frinchaboy}, {Garc{\'\i}a Perez}, {Hayden}, {Holtzman},
  {Majewski}, {M{\'e}sz{\'a}ros}, {Nidever}, {Pinsonneault}, \&
  {Shetrone}}]{mac2017}
{Mackereth}, J.~T., {Bovy}, J., {Schiavon}, R.~P., {et~al.} 2017, \mnras, 471,
  3057

\bibitem[{Majewski {et~al.}(2017)Majewski, Schiavon, Frinchaboy, Prieto,
  Barkhouser, Bizyaev, Blank, Brunner, Burton, Carrera, Chojnowski, Cunha,
  Epstein, Fitzgerald, P{\'e}rez, Hearty, Henderson, Holtzman, Johnson, Lam,
  Lawler, Maseman, M{\'e}sz{\'a}ros, Nelson, Nguyen, Nidever, Pinsonneault,
  Shetrone, Smee, Smith, Stolberg, Skrutskie, Walker, Wilson, Zasowski, Anders,
  Basu, Beland, Blanton, Bovy, Brownstein, Carlberg, Chaplin, Chiappini,
  Eisenstein, Elsworth, Feuillet, Fleming, Galbraith-Frew, Garc{\'\i}a,
  Garc{\'\i}a-Hern{\'a}ndez, Gillespie, Girardi, Gunn, Hasselquist, Hayden,
  Hekker, Ivans, Kinemuchi, Klaene, Mahadevan, Mathur, Mosser, Muna, Munn,
  Nichol, O'Connell, Parejko, Robin, Rocha-Pinto, Schultheis, Serenelli, Shane,
  Aguirre, Sobeck, Thompson, Troup, Weinberg, \& Zamora}]{Majewski:2017ip}
Majewski, S.~R., Schiavon, R.~P., Frinchaboy, P.~M., {et~al.} 2017, The
  Astronomical Journal, 154, 0

\bibitem[{{Matteucci}(2012)}]{matteucci2012}
{Matteucci}, F. 2012, {Chemical Evolution of Galaxies}

\bibitem[{{Matteucci} \& {Francois}(1989)}]{matteucci1989}
{Matteucci}, F. \& {Francois}, P. 1989, \mnras, 239, 885

\bibitem[{{Matteucci} {et~al.}(2019){Matteucci}, {Grisoni}, {Spitoni},
  {Zulianello}, {Rojas-Arriagada}, {Schultheis}, \& {Ryde}}]{matteucci2019}
{Matteucci}, F., {Grisoni}, V., {Spitoni}, E., {et~al.} 2019, \mnras, 487, 5363

\bibitem[{{Matteucci} {et~al.}(2009){Matteucci}, {Spitoni}, {Recchi}, \&
  {Valiante}}]{matteucci2009}
{Matteucci}, F., {Spitoni}, E., {Recchi}, S., \& {Valiante}, R. 2009, \aap,
  501, 531

\bibitem[{{Matteucci} {et~al.}(2020){Matteucci}, {Vasini}, {Grisoni}, \&
  {Schultheis}}]{matteucci2020}
{Matteucci}, F., {Vasini}, A., {Grisoni}, V., \& {Schultheis}, M. 2020, \mnras,
  494, 5534

\bibitem[{{McKee} {et~al.}(2015){McKee}, {Parravano}, \&
  {Hollenbach}}]{mckee2015}
{McKee}, C.~F., {Parravano}, A., \& {Hollenbach}, D.~J. 2015, \apj, 814, 13

\bibitem[{{Melioli} {et~al.}(2008){Melioli}, {Brighenti}, {D'Ercole}, \& {de
  Gouveia Dal Pino}}]{melioli2008}
{Melioli}, C., {Brighenti}, F., {D'Ercole}, A., \& {de Gouveia Dal Pino}, E.~M.
  2008, \mnras, 388, 573

\bibitem[{{Melioli} {et~al.}(2009){Melioli}, {Brighenti}, {D'Ercole}, \& {de
  Gouveia Dal Pino}}]{melioli2009}
{Melioli}, C., {Brighenti}, F., {D'Ercole}, A., \& {de Gouveia Dal Pino}, E.~M.
  2009, \mnras, 399, 1089

\bibitem[{Mikolaitis {et~al.}(2017)Mikolaitis, de~Laverny, Recio-Blanco, Hill,
  Worley, \& de~Pascale}]{Mikolaitis:2017gd}
Mikolaitis, S., de~Laverny, P., Recio-Blanco, A., {et~al.} 2017, Astronomy and
  Astrophysics, 600, A22

\bibitem[{{Minchev} {et~al.}(2011){Minchev}, {Famaey}, {Combes}, {Di Matteo},
  {Mouhcine}, \& {Wozniak}}]{minchev2011}
{Minchev}, I., {Famaey}, B., {Combes}, F., {et~al.} 2011, \aap, 527, A147

\bibitem[{{Minchev} {et~al.}(2012){Minchev}, {Famaey}, {Quillen}, {Di Matteo},
  {Combes}, {Vlaji{\'c}}, {Erwin}, \& {Bland -Hawthorn}}]{minchev2012}
{Minchev}, I., {Famaey}, B., {Quillen}, A.~C., {et~al.} 2012, \aap, 548, A126

\bibitem[{{Mor} {et~al.}(2019){Mor}, {Robin}, {Figueras}, {Roca-F{\`a}brega},
  \& {Luri}}]{mor2019}
{Mor}, R., {Robin}, A.~C., {Figueras}, F., {Roca-F{\`a}brega}, S., \& {Luri},
  X. 2019, \aap, 624, L1

\bibitem[{{Mott} {et~al.}(2013){Mott}, {Spitoni}, \& {Matteucci}}]{mott2013}
{Mott}, A., {Spitoni}, E., \& {Matteucci}, F. 2013, \mnras, 435, 2918

\bibitem[{{Nakanishi} \& {Sofue}(2003)}]{Nakanishi2003}
{Nakanishi}, H. \& {Sofue}, Y. 2003, \pasj, 55, 191

\bibitem[{{Nakanishi} \& {Sofue}(2006)}]{Nakanishi2006}
{Nakanishi}, H. \& {Sofue}, Y. 2006, \pasj, 58, 847

\bibitem[{Nidever {et~al.}(2014)Nidever, Bovy, Bird, Andrews, Hayden, Holtzman,
  Majewski, Smith, Robin, Garc{\'\i}a~P{\'e}rez, Cunha, Allende~Prieto,
  Zasowski, Schiavon, Johnson, Weinberg, Feuillet, Schneider, Shetrone, Sobeck,
  Garc{\'\i}a-Hern{\'a}ndez, Zamora, Rix, Beers, Wilson, O'Connell, Minchev,
  Chiappini, Anders, Bizyaev, Brewington, Ebelke, Frinchaboy, Ge, Kinemuchi,
  Malanushenko, Malanushenko, Marchante, M{\'e}sz{\'a}ros, Oravetz, Pan,
  Simmons, \& Skrutskie}]{Nidever:2014fj}
Nidever, D.~L., Bovy, J., Bird, J.~C., {et~al.} 2014, The Astrophysical
  Journal, 796, 38

\bibitem[{{Nissen} {et~al.}(2020){Nissen}, {Christensen-Dalsgaard},
  {Mosumgaard}, {Silva Aguirre}, {Spitoni}, \& {Verma}}]{nissen2020}
{Nissen}, P.~E., {Christensen-Dalsgaard}, J., {Mosumgaard}, J.~R., {et~al.}
  2020, \aap, 640, A81

\bibitem[{{Noguchi}(2018)}]{noguchi2018}
{Noguchi}, M. 2018, \nat, 559, 585

\bibitem[{{Palla} {et~al.}(2020){Palla}, {Matteucci}, {Spitoni}, {Vincenzo}, \&
  {Grisoni}}]{palla2020}
{Palla}, M., {Matteucci}, F., {Spitoni}, E., {Vincenzo}, F., \& {Grisoni}, V.
  2020, \mnras, 498, 1710

\bibitem[{{Philcox} {et~al.}(2018){Philcox}, {Rybizki}, \&
  {Gutcke}}]{philcox2018}
{Philcox}, O., {Rybizki}, J., \& {Gutcke}, T.~A. 2018, \apj, 861, 40

\bibitem[{{Prantzos} {et~al.}(2018){Prantzos}, {Abia}, {Limongi}, {Chieffi}, \&
  {Cristallo}}]{prantzos2018}
{Prantzos}, N., {Abia}, C., {Limongi}, M., {Chieffi}, A., \& {Cristallo}, S.
  2018, \mnras, 476, 3432

\bibitem[{{Queiroz} {et~al.}(2020){Queiroz}, {Anders}, {Chiappini},
  {Khalatyan}, {Santiago}, {Steinmetz}, {Valentini}, {Miglio}, {Bossini},
  {Barbuy}, {Minchev}, {Minniti}, {Garc{\'\i}a Hern{\'a}ndez}, {Schultheis},
  {Beaton}, {Beers}, {Bizyaev}, {Brownstein}, {Cunha},
  {Fern{\'a}ndez-Trincado}, {Frinchaboy}, {Lane}, {Majewski}, {Nataf},
  {Nitschelm}, {Pan}, {Roman-Lopes}, {Sobeck}, {Stringfellow}, \&
  {Zamora}}]{queiroz2020}
{Queiroz}, A.~B.~A., {Anders}, F., {Chiappini}, C., {et~al.} 2020, \aap, 638,
  A76

\bibitem[{{Queiroz} {et~al.}(2018){Queiroz}, {Anders}, {Santiago}, {Chiappini},
  {Steinmetz}, {Dal Ponte}, {Stassun}, {da Costa}, {Maia}, {Crestani}, {Beers},
  {Fern{\'a}ndez-Trincado}, {Garc{\'\i}a-Hern{\'a}ndez}, {Roman-Lopes}, \&
  {Zamora}}]{queiroz2018}
{Queiroz}, A.~B.~A., {Anders}, F., {Santiago}, B.~X., {et~al.} 2018, \mnras,
  476, 2556

\bibitem[{{Rana}(1991)}]{rana1991}
{Rana}, N.~C. 1991, \araa, 29, 129

\bibitem[{Recio-Blanco {et~al.}(2014)Recio-Blanco, de~Laverny, Kordopatis,
  Helmi, Hill, Gilmore, Wyse, Adibekyan, Randich, Asplund, Feltzing, Jeffries,
  Micela, Vallenari, Alfaro, Allende~Prieto, Bensby, Bragaglia, Flaccomio,
  Koposov, Korn, Lanzafame, Pancino, Smiljanic, Jackson, Lewis, Magrini,
  Morbidelli, Prisinzano, Sacco, Worley, Hourihane, Bergemann, Costado, Heiter,
  Joffre, Lardo, Lind, \& Maiorca}]{RecioBlanco:2014dd}
Recio-Blanco, A., de~Laverny, P., Kordopatis, G., {et~al.} 2014, Astronomy and
  Astrophysics, 567, A5

\bibitem[{{Renaud} {et~al.}(2020{\natexlab{a}}){Renaud}, {Agertz}, {Andersson},
  {Read}, {Ryde}, {Bensby}, {Rey}, \& {Feuillet}}]{VINTERGATANIII2020}
{Renaud}, F., {Agertz}, O., {Andersson}, E.~P., {et~al.} 2020{\natexlab{a}},
  arXiv e-prints, arXiv:2006.06012

\bibitem[{{Renaud} {et~al.}(2020{\natexlab{b}}){Renaud}, {Agertz}, {Read},
  {Ryde}, {Andersson}, {Bensby}, {Rey}, \& {Feuillet}}]{VINTERGATANII2020}
{Renaud}, F., {Agertz}, O., {Read}, J.~I., {et~al.} 2020{\natexlab{b}}, arXiv
  e-prints, arXiv:2006.06011

\bibitem[{Rojas-Arriagada {et~al.}(2017)Rojas-Arriagada, Recio-Blanco,
  de~Laverny, Mikolaitis, Matteucci, Spitoni, Schultheis, Hayden, Hill,
  Zoccali, Minniti, Gonzalez, Gilmore, Randich, Feltzing, Alfaro, Babusiaux,
  Bensby, Bragaglia, Flaccomio, Koposov, Pancino, Bayo, Carraro, Casey,
  Costado, Damiani, Donati, Franciosini, Hourihane, Jofre, Lardo, Lewis, Lind,
  Magrini, Morbidelli, Sacco, Worley, \& Zaggia}]{RojasArriagada:2017ka}
Rojas-Arriagada, A., Recio-Blanco, A., de~Laverny, P., {et~al.} 2017, Astronomy
  and Astrophysics, 601, A140

\bibitem[{Rojas-Arriagada {et~al.}(2016)Rojas-Arriagada, Recio-Blanco,
  de~Laverny, Schultheis, Guiglion, Mikolaitis, Kordopatis, Hill, Gilmore,
  Randich, Alfaro, Bensby, Koposov, Costado, Franciosini, Hourihane, Jofre,
  Lardo, Lewis, Lind, Magrini, Monaco, Morbidelli, Sacco, Worley, Zaggia, \&
  Chiappini}]{RojasArriagada:2016eq}
Rojas-Arriagada, A., Recio-Blanco, A., de~Laverny, P., {et~al.} 2016, Astronomy
  and Astrophysics, 586, A39

\bibitem[{{Rojas-Arriagada} {et~al.}(2020){Rojas-Arriagada}, {Zasowski},
  {Schultheis}, {Zoccali}, {Hasselquist}, {Chiappini}, {Cohen}, {Cunha},
  {Fern{\'a}ndez-Trincado}, {Fragkoudi}, {Garc{\'\i}a-Hern{\'a}ndez},
  {Geisler}, {Gran}, {Lian}, {Majewski}, {Minniti}, {Monachesi}, {Nitschelm},
  \& {Queiroz}}]{rojas2020}
{Rojas-Arriagada}, A., {Zasowski}, G., {Schultheis}, M., {et~al.} 2020, \mnras,
  499, 1037

\bibitem[{{Romano} {et~al.}(2010){Romano}, {Karakas}, {Tosi}, \&
  {Matteucci}}]{romano2010}
{Romano}, D., {Karakas}, A.~I., {Tosi}, M., \& {Matteucci}, F. 2010, \aap, 522,
  A32

\bibitem[{{Ro{\v{s}}kar} {et~al.}(2008){Ro{\v{s}}kar}, {Debattista}, {Stinson},
  {Quinn}, {Kaufmann}, \& {Wadsley}}]{roskar2008}
{Ro{\v{s}}kar}, R., {Debattista}, V.~P., {Stinson}, G.~S., {et~al.} 2008,
  \apjl, 675, L65

\bibitem[{{Rybizki} {et~al.}(2017){Rybizki}, {Just}, \& {Rix}}]{rybi2017}
{Rybizki}, J., {Just}, A., \& {Rix}, H.-W. 2017, \aap, 605, A59

\bibitem[{{Salaris} {et~al.}(1993){Salaris}, {Chieffi}, \&
  {Straniero}}]{salaris1993}
{Salaris}, M., {Chieffi}, A., \& {Straniero}, O. 1993, \apj, 414, 580

\bibitem[{{Scalo}(1986)}]{scalo1986}
{Scalo}, J.~M. 1986, \fcp, 11, 1

\bibitem[{{Sch{\"o}nrich} \& {Binney}(2009)}]{schoenrich2009MNRAS}
{Sch{\"o}nrich}, R. \& {Binney}, J. 2009, \mnras, 396, 203

\bibitem[{{Sch{\"o}nrich} \& {McMillan}(2017)}]{schoenrich2017}
{Sch{\"o}nrich}, R. \& {McMillan}, P.~J. 2017, \mnras, 467, 1154

\bibitem[{{Sellwood} \& {Binney}(2002)}]{sellwood2002}
{Sellwood}, J.~A. \& {Binney}, J.~J. 2002, \mnras, 336, 785

\bibitem[{{Sharma} {et~al.}(2020){Sharma}, {Hayden}, \&
  {Bland-Hawthorn}}]{sharma2020}
{Sharma}, S., {Hayden}, M.~R., \& {Bland-Hawthorn}, J. 2020, arXiv e-prints,
  arXiv:2005.03646

\bibitem[{Silva~Aguirre {et~al.}(2018)Silva~Aguirre, Bojsen-Hansen, Slumstrup,
  Casagrande, Kawata, Ciuc{\v a}, Handberg, Lund, Mosumgaard, Huber, Johnson,
  Pinsonneault, Serenelli, Stello, Tayar, Bird, Cassisi, Hon, Martig, Nissen,
  Rix, Sch{\"o}nrich, Sahlholdt, Trick, \& Yu}]{victor2018}
Silva~Aguirre, V., Bojsen-Hansen, M., Slumstrup, D., {et~al.} 2018, Monthly
  Notices of the Royal Astronomical Society, 475, 5487

\bibitem[{Snaith {et~al.}(2015)Snaith, Haywood, Di~Matteo, Lehnert, Combes,
  Katz, \& Gomez}]{snaith2015}
Snaith, O., Haywood, M., Di~Matteo, P., {et~al.} 2015, Astronomy and
  Astrophysics, 578, A87

\bibitem[{{Spitoni} {et~al.}(2019{\natexlab{a}}){Spitoni}, {Cescutti},
  {Minchev}, {Matteucci}, {Silva Aguirre}, {Martig}, {Bono}, \&
  {Chiappini}}]{spitoni2D2018}
{Spitoni}, E., {Cescutti}, G., {Minchev}, I., {et~al.} 2019{\natexlab{a}},
  \aap, 628, A38

\bibitem[{{Spitoni} {et~al.}(2017){Spitoni}, {Gioannini}, \&
  {Matteucci}}]{spitoni2017}
{Spitoni}, E., {Gioannini}, L., \& {Matteucci}, F. 2017, \aap, 605, A38

\bibitem[{{Spitoni} {et~al.}(2009){Spitoni}, {Matteucci}, {Recchi}, {Cescutti},
  \& {Pipino}}]{spitoni2009}
{Spitoni}, E., {Matteucci}, F., {Recchi}, S., {Cescutti}, G., \& {Pipino}, A.
  2009, \aap, 504, 87

\bibitem[{{Spitoni} {et~al.}(2014){Spitoni}, {Matteucci}, \&
  {Sozzetti}}]{spitoni2014}
{Spitoni}, E., {Matteucci}, F., \& {Sozzetti}, A. 2014, \mnras, 440, 2588

\bibitem[{{Spitoni} {et~al.}(2008){Spitoni}, {Recchi}, \&
  {Matteucci}}]{spitoni2008}
{Spitoni}, E., {Recchi}, S., \& {Matteucci}, F. 2008, \aap, 484, 743

\bibitem[{{Spitoni} {et~al.}(2015){Spitoni}, {Romano}, {Matteucci}, \&
  {Ciotti}}]{spitoni2015}
{Spitoni}, E., {Romano}, D., {Matteucci}, F., \& {Ciotti}, L. 2015, \apj, 802,
  129

\bibitem[{{Spitoni} {et~al.}(2019{\natexlab{b}}){Spitoni}, {Silva Aguirre},
  {Matteucci}, {Calura}, \& {Grisoni}}]{spitoni2019}
{Spitoni}, E., {Silva Aguirre}, V., {Matteucci}, F., {Calura}, F., \&
  {Grisoni}, V. 2019{\natexlab{b}}, \aap, 623, A60

\bibitem[{{Spitoni} {et~al.}(2020){Spitoni}, {Verma}, {Silva Aguirre}, \&
  {Calura}}]{spitoni2020}
{Spitoni}, E., {Verma}, K., {Silva Aguirre}, V., \& {Calura}, F. 2020, \aap,
  635, A58

\bibitem[{{Tayar} {et~al.}(2017){Tayar}, {Somers}, {Pinsonneault}, {Stello},
  {Mints}, {Johnson}, {Zamora}, {Garc{\'\i}a-Hern{\'a}ndez}, {Maraston},
  {Serenelli}, {Allende Prieto}, {Bastien}, {Basu}, {Bird}, {Cohen}, {Cunha},
  {Elsworth}, {Garc{\'\i}a}, {Girardi}, {Hekker}, {Holtzman}, {Huber},
  {Mathur}, {M{\'e}sz{\'a}ros}, {Mosser}, {Shetrone}, {Silva Aguirre},
  {Stassun}, {Stringfellow}, {Zasowski}, \& {Roman-Lopes}}]{tayar2017}
{Tayar}, J., {Somers}, G., {Pinsonneault}, M.~H., {et~al.} 2017, \apj, 840, 17

\bibitem[{{Vincenzo} \& {Kobayashi}(2020)}]{vincenzo2020}
{Vincenzo}, F. \& {Kobayashi}, C. 2020, \mnras, 496, 80

\bibitem[{{Vincenzo} {et~al.}(2019){Vincenzo}, {Spitoni}, {Calura},
  {Matteucci}, {Silva Aguirre}, {Miglio}, \& {Cescutti}}]{vincenzo2019}
{Vincenzo}, F., {Spitoni}, E., {Calura}, F., {et~al.} 2019, \mnras, L74

\bibitem[{{Weinberg} {et~al.}(2019){Weinberg}, {Holtzman}, {Hasselquist},
  {Bird}, {Johnson}, {Shetrone}, {Sobeck}, {Allende Prieto}, {Bizyaev},
  {Carrera}, {Cohen}, {Cunha}, {Ebelke}, {Fernandez-Trincado},
  {Garc{\'\i}a-Hern{\'a}ndez}, {Hayes}, {J{\"o}nsson}, {Lane}, {Majewski},
  {Malanushenko}, {M{\'e}sz{\'a}ros}, {Nidever}, {Nitschelm}, {Pan}, {Rix},
  {Rybizki}, {Schiavon}, {Schneider}, {Wilson}, \& {Zamora}}]{weinberg2019}
{Weinberg}, D.~H., {Holtzman}, J.~A., {Hasselquist}, S., {et~al.} 2019, \apj,
  874, 102

\bibitem[{{Woosley} \& {Weaver}(1995)}]{WW1995}
{Woosley}, S.~E. \& {Weaver}, T.~A. 1995, \apjs, 101, 181

\bibitem[{{Wyse} \& {Silk}(1989)}]{wyse1989}
{Wyse}, R. F.~G. \& {Silk}, J. 1989, \apj, 339, 700

\end{thebibliography}

\end{document}